\documentclass{emulateapj}

\usepackage{lscape}
\usepackage{apjfonts}
\newcommand \msun{\hbox{$\hbox{M}_{\odot}$}} 
  
\slugcomment{Accepted for publication in The Astronomical Journal (January 2005 issue)}
\shorttitle{HI Parkes Northern ZOA Survey}
\shortauthors{DONLEY ET~AL}

\begin{document}

\title{The HI Parkes Zone of Avoidance Survey: the Northern Extension}

\author{
J. L. Donley,\altaffilmark{1,2}
L. Staveley-Smith,\altaffilmark{1}
R. C. Kraan-Korteweg,\altaffilmark{3}
J. M. Islas-Islas,\altaffilmark{3}
A. Schr\"{o}der,\altaffilmark{4}
P. A. Henning,\altaffilmark{5}
B. Koribalski,\altaffilmark{1}
S. Mader,\altaffilmark{1} 
and I. Stewart \altaffilmark{4}}

\altaffiltext{1}{Australia Telescope National Facility, CSIRO, P.O. Box 76, Epping, NSW 1710, Australia; Lister.Staveley-Smith@csiro.au, Baerbel.Koribalski@csiro.au, Stacy.Mader@csiro.au}
\altaffiltext{2}{Steward Observatory, University of Arizona, 933 North Cherry Avenue, Tucson, AZ 85721; jdonley@as.arizona.edu}
\altaffiltext{3}{Departamento de Astronom\'\i{a}, Universidad de Guanajuato, Apdo. Postal 144, Guanajuato, GTO 36000, Mexico; kraan@astro.ugto.mx, jmislas@astro.ugto.mx}
\altaffiltext{4}{Department of Physics and Astronomy, University of Leicester, Leicester LE1 7RH, UK; acs@star.le.ac.uk, ims@star.le.ac.uk}
\altaffiltext{5}{Institute for Astrophysics, University of New Mexico, 800 Yale Boulevard, NE, Albuquerque, NM 87131; henning@cosmos.phys.unm.edu}


\begin{abstract}

We present the results of the northern extension of the HI Parkes Zone
of Avoidance Survey, a blind HI survey utilizing the multibeam
receiver on the Parkes 64-m telescope.  In the two regions studied
here, $l=36^{\circ}$ to $52^{\circ}$ and $l=196^{\circ}$ to
$212^{\circ}$, $|b|\le5^{\circ}$, we have detected 77 HI galaxies,
twenty of which have been previously detected in HI.  The survey has a
median rms noise of 6.0~mJy~beam$^{-1}$ and is complete to a mean flux
density of 22~mJy.  We have searched for multiwavelength counterparts
to the 77 galaxies detected here: 19, 27, and 11 have a likely
optical, 2MASS, and IRAS cataloged counterpart, respectively.  A
further 16 galaxies have likely visible counterparts on the Digitized
Sky Survey. The detection of these 77 galaxies allows a closer
inspection of the large-scale structures in these regions.  We see
several filaments crossing the Galactic plane, one of which appears to
be the continuation of a sine-wave like feature that can be traced
across the whole southern sky. An analysis of the HI mass function
suggests that the regions studied here may be underdense.  One
particularly noteworthy galaxy is HIZOA J0630+08 ($l,b$ =
203$^{\circ}$, --0.9$^{\circ}$) with a velocity of $367\pm
1$~km~s$^{-1}$. We suggest that it belongs to the nearby Orion Group
which includes a small number of dwarf galaxies. The newly detected
galaxies improve our understanding of the properties of several voids,
such as the Orion, Gemini, and Canis Major Voids.

\end{abstract}



\section{Introduction}

The obscuring effects of dust and high stellar density in the Milky
Way have historically prevented the detection of galaxies that lie
behind the disk of our Galaxy. This zone of avoidance (ZOA) typically
comprises more than 20\% of the optical extragalactic sky and nearly
10\% of the infrared extragalactic sky, although recent surveys have
begun to lessen its extent. The presence of the ZOA does not affect
extragalactic population studies, as galaxies in the highly obscured
ZOA should not differ from those detected in regions of low
obscuration.  A complete census of galaxies across the Galactic plane
is necessary, however, to fully understand the dynamics of the Local
Group as well as the large-scale structure of the Local Universe.

Many early ZOA studies focused on finding a highly obscured, nearby,
massive galaxy.  Several such galaxies have been found, including
Maffei~1~and~2 (Maffei 1968), Circinus (Freeman et al. 1977), and
Dwingeloo 1 (Kraan-Korteweg et al. 1994).  While the IC~342/Maffei
group and other nearby massive galaxies largely define the motion of
the Local Group, the direction of the Local Group acceleration is
strongly affected by the full ZOA mass distribution out to
6000~km~s$^{-1}$ (Kolatt, Dekel, \& Lahav 1995), and possibly beyond.

Several attempts have been made in recent years to sample the ZOA more
completely. Optical galaxy candidates have been identified by searches
of the Palomar Observatory Sky Survey (POSS) prints and Schmidt
atlases (see Kraan-Korteweg \& Lahav 2000 for a review). Highly
obscured galaxies have also been detected in the near-infrared (NIR)
and far-infrared by the Deep Near-Infrared Survey of the Southern Sky
(DENIS; Schr\"{o}der, Kraan-Korteweg, \& Mamon 1999), the Two Micron
All Sky Survey (2MASS; Jarrett et al. 2000), and the Infrared
Astronomical Satellite Sky Survey (IRAS, e.g. Nakanishi et al. 1997).

While useful in reducing the size of the ZOA, optical and infrared
surveys are ultimately limited by Galactic extinction and source
confusion. The most thorough optical searches are complete to an
apparent diameter of $D=14^{\prime\prime}$ and a magnitude of
B$_{25}=18^{\rm m}\!\!.5$ at extinctions of $A_{\rm B} \le 3^{\rm
m}\!\!.0$ (Kraan-Korteweg 2000, Woudt \& Kraan-Korteweg 2001).  The
DIRBE/\textit{IRAS} 100~$\mu$m extinction maps (Schlegel, Finkbeiner,
\& Davis 1998) directly measure the dust column density and indicate
that $A_{\rm B} \ge 3^{\rm m}\!\!.0$ for 9.5\% of the sky, although
the data in the Galactic plane are not yet well-calibrated. The
completeness of infrared surveys of the ZOA is affected primarily by
source confusion. Results from the DENIS survey indicate that NIR
surveys can easily detect galaxies down to $A_{\rm B} \sim 10$~mag,
and the brighter ones far beyond (Schr\"{o}der et al. 1999).  The
2MASS survey is expected to be complete to $13^{\rm m}\!\!.5$ at
2.2~$\mu$m for the majority of the sky, but for $5^{\circ} \le |b| \le
20^{\circ}$, this completeness drops to $12^{\rm m}\!\!.1$ (Jarrett et
al. 2000).

HI emission is not affected by the dust in the Milky Way.
Consequently, blind HI surveys provide a unique means by which to
identify gas-rich ZOA galaxies. The first blind ZOA HI survey was
conducted by Henning (1992) using the Green Bank 300-ft
telescope. More recent blind HI surveys include the Arecibo Dual Beam
Survey (ADBS, Rosenberg \& Schneider 2000) and the Arecibo HI Strip
Survey (AHISS, Zwaan et al. 1997), which intersect with but do not
focus on the ZOA. The first systematic HI survey to focus on the
optically most opaque part of the ZOA is the Dwingeloo Obscured
Galaxies Survey (Henning et al. 1998, Rivers 2000) which sampled the
northern ZOA ($30^{\circ} \le l \le 220^{\circ}$, $|b| \le
5^{\circ}\!\!.25$) to velocities of 4000~km~s$^{-1}$ in the Local
Group standard of rest with an rms of 40~mJy~beam$^{-1}$. Each of
these surveys has contributed to our understanding of gas-rich
galaxies and the structure of the local Universe, but all have been
limited in either sensitivity or sky coverage.

The 21-cm multibeam receiver (Staveley-Smith et al. 1996) on the
Parkes\footnote{The Parkes telescope is part of the Australia
Telescope which is funded by the Commonwealth of Australia for
operation as a National Facility managed by CSIRO.} 64-m telescope is
the first instrument capable of conducting both sensitive \textit{and}
large-area blind HI surveys in a reasonable time.  Two multibeam
surveys commenced in 1997. The HI Parkes All-Sky Survey (HIPASS, see
Meyer et al. 2004) is a blind HI survey of the sky south of $\delta =
25^{\circ}$ and has an rms noise of 13~mJy~beam$^{-1}$.  The HI Parkes
Zone of Avoidance Survey is a more sensitive (rms = 6 mJy~beam$^{-1}$)
survey of the region of the Milky Way accessible from Parkes ($l=
196^{\circ}$ to $52^{\circ}$, $|b| < 5^{\circ}$).  An analysis of
shallower data for the southern ZOA (rms = 15~mJy~beam$^{-1}$, $l=
212^{\circ}$ to $36^{\circ}$) led to the discovery of 110 galaxies, 67
of which were previously unknown (Henning et al. 2000). Here we report
on the results for the complementary region north of Dec. $0^{\circ}$
($l=36^{\circ}$ to $52^{\circ}$ and $l=196^{\circ}$ to $212^{\circ}$).
The results reported here are, however, at full survey sensitivity
(rms = 6~mJy~beam$^{-1}$).  Although this region overlaps with the
Dwingeloo survey, the Parkes survey is much more sensitive and
represents the first probe, beyond a few Mpc, of this region of the
Universe, which lies near both the Local and Microscopium Voids.  Data
of similar depth for the remainder of the ZOA ($l= 212^{\circ}$ to
$36^{\circ}$) will be reported elsewhere.

In \S 2, we describe the observations and the method by which the data
were reduced.  The galaxy search method, measurement of HI parameters,
search results, and survey completeness are discussed in \S 3. In \S
4, we describe the results of a search for multiwavelength
counterparts to the survey galaxies.  The large-scale structures
revealed by the survey are discussed in \S 5, and the mass function of
the galaxies is discussed in \S 6. Throughout the paper, we assume
$H_0= 75$~km~s$^{-1}$~Mpc$^{-1}$.


\section{Observations and Data Reduction}

The observations for the northern extension of the HI Parkes ZOA
survey began in 2000 October and were completed in 2002 May. The data
were taken at the Parkes 64-m telescope using the 21-cm multibeam
receiver, which consists of an array of 13 beams each with two
orthogonal linear polarizations and an average beamwidth of
$14^{\prime}\!\!.3$ (FWHM).  The multibeam correlator has a bandwidth
of 64~MHz and covers the velocity range $-1200 < cz
<12700$~km~s$^{-1}$.  The channel spacing is 13.2~km~s$^{-1}$ and the
average system temperature is 20K.

The northern ZOA region referred to in this paper lies at declinations
$\ga 0^{\circ}$ and is divided into four fields, each approximately
$8^{\circ}\times10^{\circ}$, as shown in Figure 1. Each field was
scanned 425 times, with a single $8^{\circ}$ scan covering a region
$1.7^{\circ}$ wide in eight minutes.  The central beam for all scans
lies along a line of constant Galactic latitude, with each of the 425
scans being separated by 1\farcm4 in latitude.  At the midpoint of
each scan, the feed angle is $15^{\circ}$ with respect to the scan
direction so that all 13 beams lie at different latitudes in a
minimally redundant fashion. The total integration time is
2100~s~beam$^{-1}$.  The median rms noise of the four fields after
Hanning smoothing is 6.0~mJy~beam$^{-1}$.

\begin{figure*}[t!]
\centering
\epsscale{0.9}
\plotone{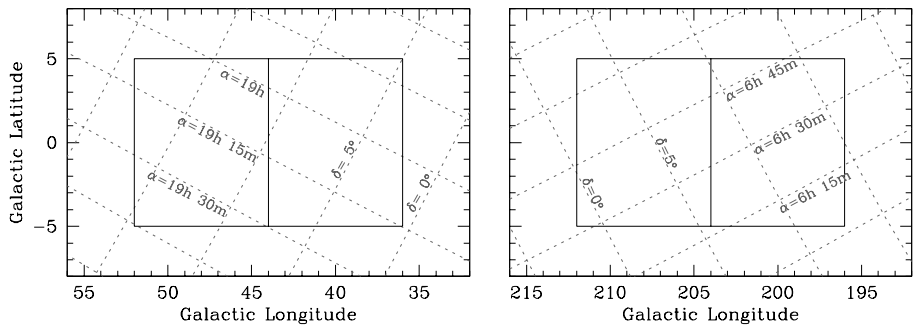}
\caption{The survey region consists of four fields, each approximately 
$8^{\circ} \times 10^{\circ}$.}
\end{figure*}

The data were bandpass-corrected, Doppler-corrected, and calibrated
using AIPS++ {\sc livedata} (Barnes et al. 2001). Gridding was
performed by {\sc gridzilla}, using a top-hat median gridding
algorithm with a top-hat radius of 6$^{\prime}$ (Barnes et
al. 2001). The resulting pixel and beam sizes are $4^{\prime}\times
4^{\prime}$ and $15^{\prime}\!\!.5$, respectively. Spectral ringing
associated with Galactic HI emission and the baseline ripple
associated with strong continuum emission were respectively removed or
suppressed using Hanning smoothing and the 'scaled template method'
(see Barnes et al. 2001). Hanning smoothing increases the velocity
resolution to 27~km~s$^{-1}$.

Several sources of extraneous signal remained after standard
processing. Residual baseline ripples caused by continuum radio
sources are the most prominent. Hydrogen recombination lines at
1424.734 MHz (H166$\alpha$), 1399.368 MHz (H167$\alpha$), and 1374.601
MHz (H168$\alpha$) appear at velocities of cz=-911~km~s$^{-1}$,
4507~km~s$^{-1}$, and 9990~km~s$^{-1}$, respectively. Interference
from the L3 beacon of the GPS near 1380 MHz (8778~km~s$^{-1}$) is also
present. These sources of interference can generally be identified by
their characteristic frequencies and widths. In the case of
recombination lines, they are accompanied by continuum emission from
their underlying HII regions.


\section{Galaxy Search}

The four Hanning-smoothed data cubes were visualized using {\sc kview}
in the software package {\sc karma} (Gooch 1995) and were searched by
eye; the strong residual baseline ripples prevented the use of
automatic galaxy-finding software. The galaxy search was conducted
over the full velocity range, $-1200 < cz < 12700$~km~s$^{-1}$,
although confusion from Galactic emission prevented a complete survey
of the region with $|v| < 250$~km~s$^{-1}$.  All possible planes,
RA-Dec, RA-velocity, and Dec-velocity were examined by three
independent parties and a final galaxy list was decided upon by a
fourth author.

\subsection{Determination of HI Properties} 

The coordinates and spectral characteristics of each galaxy were
measured using the program {\sc mbspect} in the software package {\sc
miriad} (Sault, Teuben, \& Wright 1995). The galaxy positions were
determined by spatially fitting a Gaussian to the velocity-integrated
emission of each galaxy. We subsequently used this position in the
spectral analysis, where the weighted sum of the emission in each
velocity plane was used to form the spectral profile. The total flux
was measured by integrating the flux over the galaxy profile. The
systemic velocity was taken to be the midpoint of the profile at 50\%
of the peak flux. The widths at 20\% and 50\% of the peak flux density
were measured using a width-maximizing technique. Due to the large
Hanning-smoothed velocity resolution of 27~km~s$^{-1}$, the velocity
widths we measure are broader than the intrinsic widths of the
galaxies. The velocity widths at the 20\% and 50\% peak flux levels
were therefore corrected by 21 and 14~km~s$^{-1}$, respectively (see
Henning et al. 2000). If the HI emission showed no evidence of
extension or confusion, we analyzed a region 28$^{\prime}$ around the
galaxy. For seven of the galaxies, a narrower box of width
12$^{\prime}$ or 20$^{\prime}$ was used to isolate the galaxy emission
from that arising from a neighboring galaxy, which may have led to an
underestimation of the total flux. While no galaxies in our survey are
clearly extended, we did identify one galaxy that is possibly extended
or confused, J1912+02. The spectral fitting for this galaxy was
conducted by summing the emission over a
$36^{\prime}\times36^{\prime}$ box.  The position was fit over a
smaller $12^{\prime}\times12^{\prime}$ box.

Where possible, a first-order baseline was fit to each profile. For
approximately half of the galaxies, however, baseline ripples made a
first-order fit unrealistic; for these galaxies, we used the lowest
order baseline that gave a reasonable fit.  The distance to each
galaxy, $D = v_{\rm LG}/\rm{H_{\rm o}}$, was calculated using its
velocity in the Local Group standard of rest, $v_{\rm LG} = v_{\rm
hel} + 300$~sin$l$~cos$b$.  The HI mass was determined using $M_{\rm
HI} [\msun] = 2.36\times10^{5} D^{2} F_{\rm HI}$, where $D$ is the
distance in Mpc and $F_{\rm HI}$ is the integrated HI flux in
Jy~km~s$^{-1}$.

To estimate the errors on the HI parameters, we conducted a set of
simulations. Each galaxy profile was first smoothed using a
Savitzky-Golay smoothing filter (Press et al. 1992, Sec. 14.8). This
least-squares polynomial filter reduces noise while better retaining
the higher-order characteristics of the spectrum than traditional
smoothing techniques.  Fifty simulated spectra were then created for
each galaxy by adding random Poisson noise to the smoothed galaxy
spectrum; the rms of this random noise was set equal to that of the
original galaxy spectrum. The errors on the HI parameters were then
taken to be the median absolute offsets between the parameters of the
50 simulated spectra, as measured by {\sc mbspect}, and those of the
smoothed galaxy profile.  The median errors on the peak and integrated
flux for the entire sample of galaxies are approximately 9\% and 4\%,
respectively. The median error on the systemic velocity is
3~km~s$^{-1}$ and the median errors on the 50\% and 20\% velocity
widths are 8~km~s$^{-1}$ and 10~km~s$^{-1}$, respectively. We note
that the error we quote for the integrated flux does not take into
account the uncertainty due to baseline subtraction or the uncertainty
in calibration; the total errors on the integrated flux are likely to
be $\sim$ 10-15\%.  We further discuss this method and the dependence
of the estimated errors on S/N in Appendix A.

\subsection{Results}

77 definite or probable galaxies were detected, 35 in the
$l=36^{\circ}$ to $52^{\circ}$ region and 42 in the $l=196^{\circ}$ to
$212^{\circ}$ region. Because adjacent data cubes extend slightly
beyond the $8^{\circ}\times10^{\circ}$ boundaries and overlap in
Galactic longitude, 5 galaxies were initially detected twice. Of
these, the detection located within the well defined
$8^{\circ}\times10^{\circ}$ region of a cube was chosen for the final
list. The HI properties of the 77 galaxies are presented in Table 1.
The HI spectra of a subset of the sample are shown in Figure 2; the
complete set of spectra is available in the electronic edition of the
Journal.  The set of dotted lines nearest the galaxy spectrum defines
the region in which the galaxy profile was measured.  A second set of
more widely spaced dotted lines indicate the region over which the
baseline was fit. If no second set is present, the baseline fit was
applied to the entire region displayed. All velocities are given in
the optical convention, $v=cz$.

For several galaxies, a highly variable baseline prevents a convincing
spectrum from being attained although the galaxy is clearly present in
the data cube. As an example, consider J0623+14, a galaxy that was
detected despite its superposition on one of the most extreme ripples
that remain in the survey data. Coincidentally, J0623+14 has
previously been observed by Arecibo in an HI follow-up survey of
Weinberger galaxies (Pantoja et al. 1994).  Despite the extremely poor
baseline we observe, our HI parameters are in satisfactory agreement
with the Arecibo HI parameters.  The Arecibo flux is higher than what
we detect by 14\% and the peak flux density is lower by 22\%. The most
significant difference between our HI properties and those derived
from the Arecibo observation is the 50\% velocity width; we measure a
value of 286~km~s$^{-1}$ whereas Pantoja et al. measure
411~km~s$^{-1}$.

\begin{figure*}[p]
\centering
\epsscale{1.2}
\plotone{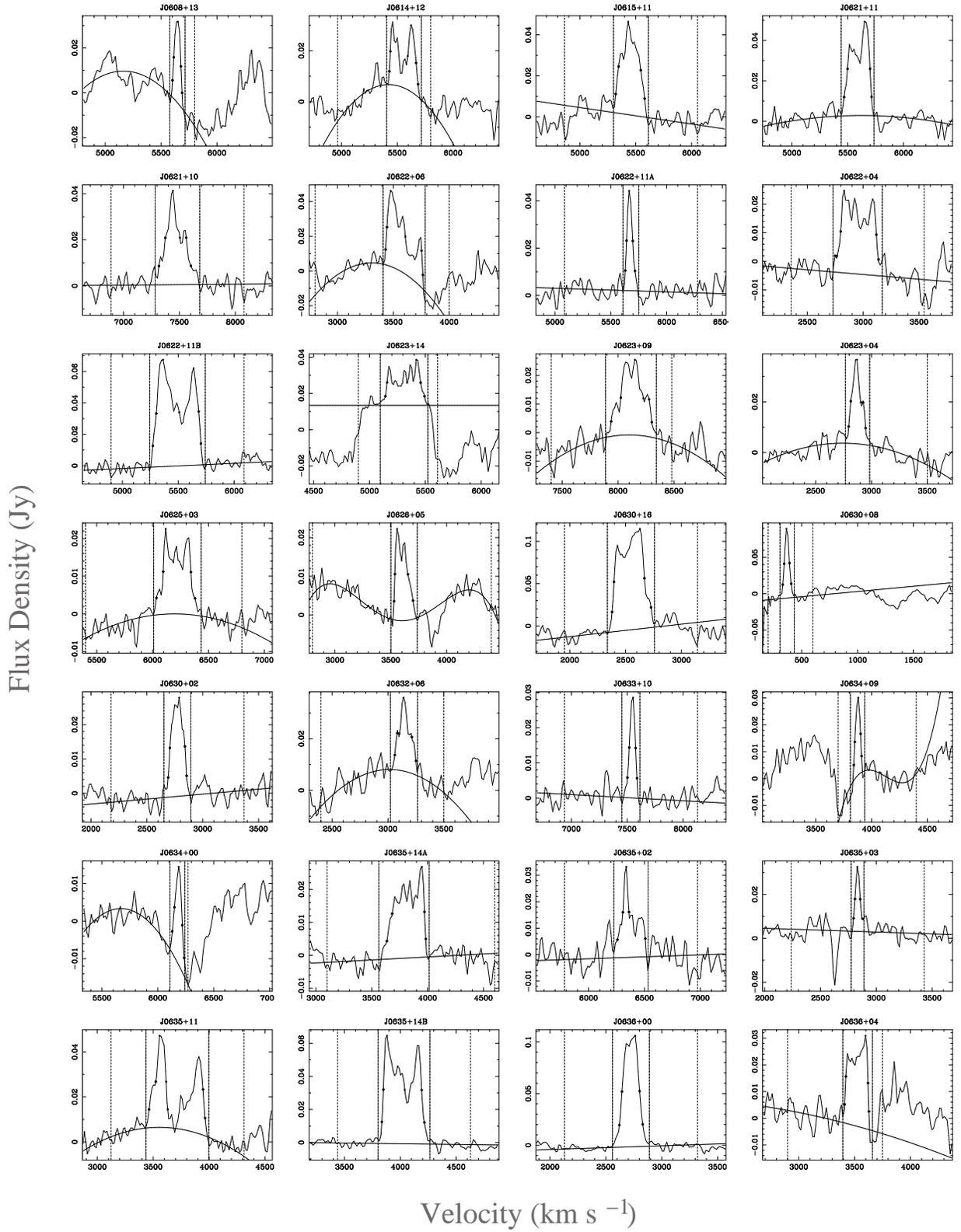}
\caption{HI spectra of a subset of the survey galaxies. The spectra
of all 77 galaxies are available in the electronic edition of the
Journal.  The inner dotted lines indicate the range in velocity over
which the galaxy profile was measured. The outer dotted lines indicate
the region over which the baseline was fit.}
\end{figure*}

The identification of confused and marginally extended galaxies in our
sample is difficult. We expect most normal galaxies to be unresolved
in the Parkes beam at the typical distances in this survey, although
confusion in the Parkes beam is not unlikely. The HI profile is the
best indicator of a confused signal, although a profile can be
intrinsically asymmetric, and minor contributions to the HI profile
often do not disturb the profile greatly. A possible confusion in the
detection can be confirmed by the relative positions and distances of
the potential optical and IR counterparts, although background
galaxies and HI-poor counterparts make this difficult, as does the
high extinction in the ZOA.  The unambiguous identification of
confused galaxies can therefore only be made by high-resolution
follow-up observations.

The mass distribution of the survey galaxies is shown in Figure 3. The
HI masses range from $1.0\times 10^{7}~\msun$ to $3.6\times
10^{10}~\msun$, with a mean mass of $4.3\times 10^9~\msun$. The
distributions of velocities in the Local Group (solid line) and
heliocentric (dotted line) standards of rest are shown in Figure 4.
By using the Local Group standard of rest, we remove the effect of the
Milky Way's velocity within the Local Group from our velocity and
distance determinations.  As we are interested in large-scale
structures outside of the Local Group, we feel that this is the more
appropriate standard of rest.  Both the velocities in the Local Group
and heliocentric standards of rest have been plotted in Figure 4,
however, for comparison.  Distinct peaks can be seen at Local Group
standard of rest velocities near 3250~km~s$^{-1}$, 4750~km~s$^{-1}$,
and 6750~km~s$^{-1}$ in the $l=36^{\circ}$ to $52^{\circ}$ region
which seem to be due to filaments crossing the Galactic plane at those
velocities (see also Fig. 10).  The first of these peaks seems to be
due to the boundary of the Local Void.  The velocity distribution for
the $l=196^{\circ}$ to $212^{\circ}$ region seems smoother overall,
except for the curious gap around 4500~km~s$^{-1}$. As will be
discussed in \S5, this structure may be due to the wall of the Canis
Major Void.

\begin{figure}
\epsscale{0.9}
\plotone{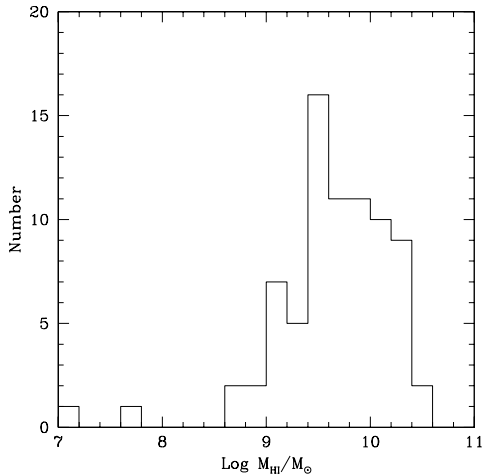}
\caption{Mass distribution of galaxies detected in the survey. The mean
mass is $4.3 \times 10^{9}$ \msun.}
\end{figure}

\subsection{Nearby Galaxies}

Of the 77 galaxies detected in this survey, three have systemic
velocities of $cz < 700$~km~s$^{-1}$. The newly detected galaxy HIZOA
J0630+08 ($l,b$ = 203.1\degr,--0.9\degr) has a velocity of only $367
\pm 1$km~s$^{-1}$ (see Table 1) and could be a high velocity cloud,
though its isolation and relative compactness suggest it is more
likely a nearby dwarf galaxy.  The Galactic extinction towards this
galaxy is $A_{\rm B} \sim 3$ mag (Schlegel et al. 1998). HIZOA
J0630+08 may be part of the Orion Group (Giovanelli \& Haynes 1981)
whose known members are the Orion Dwarf ($l,b$ = 200.6\degr,
--12.3\degr), the irregular galaxy 0554+07 ($l,b$ = 200.0\degr,
--8.4\degr), the dwarf irregular galaxy CGCG~422--003 ($l,b$ =
198.8\degr, --12.2\degr) and UGC~03303 ($l,b$ = 198.6,
--16.9\degr). Their velocities are cz = 365, 411, 441 and 521
km~s$^{-1}$, respectively (Giovanelli 1979, Giovanelli \& Haynes 1981,
Michel \& Huchra 1988, Thuan \& Seitzer 1979).  Karachentsev \&
Musella (1996) give distances of $6.4 \pm 2.2$ Mpc (Orion Dwarf) and
$5.5 \pm 1.9$ Mpc (0554+07) assuming Galactic extinctions of 2.7 and
2.9 mag, respectively. At a distance of 6 Mpc the HI mass of HIZOA
J0630+08 would be $4\times10^{7}$ \msun. Further investigations of the
Orion Group and surroundings are under way.

\begin{figure}[t]
\epsscale{1.2}
\plotone{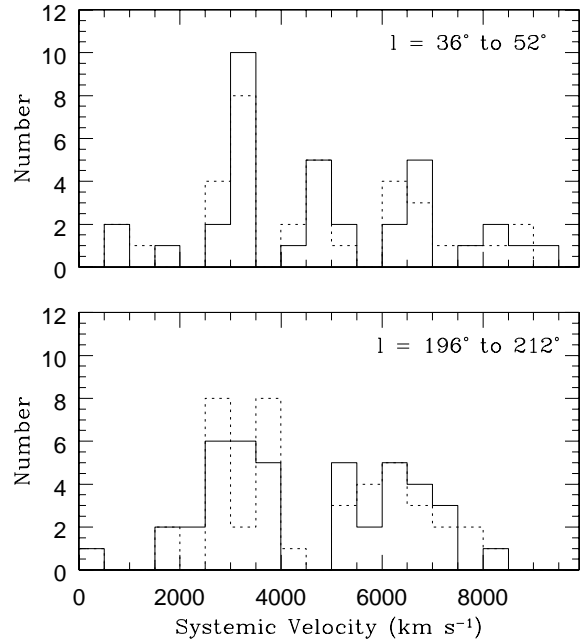}
\caption{Velocity distribution for galaxies in the $l=36^{\circ}$ to
 $52^{\circ}$ and $l=196^{\circ}$ to $212^{\circ}$ regions. The
 velocity distributions in the Local Group and heliocentric standards
 of rest are given by a solid line and a dotted line, respectively.}
\end{figure}

In addition to HIZOA J0630+08, we detect two other nearby galaxies,
HIZOA J1940+11 and HIZOA J1914+10, with systemic velocities of
580~km~s$^{-1}$ and 654~km~s$^{-1}$, respectively.  For HIZOA J1914+10
we derive an HI mass of $M_{\rm HI} = 6.5 \times 10^8$
M$_{\odot}$. This galaxy was previously detected in the Dwingeloo
Obscured Galaxies Survey and is known as Dw044.8--0.5 (Henning et
al. 1998, Rivers 2000). The Galactic extinction in this direction is
extremely high, preventing detection of optical or infrared
counterparts.

For HIZOA J1940+11, a newly cataloged galaxy, we derive an HI mass of
only $M_{\rm HI} = 4.1 \times 10^7$ M$_{\odot}$.  Confirmation of this
HI detection would be desirable as it is very weak and found at the
edge of our survey field. The Galactic extinction in this direction is
$A_{\rm B} \sim 2$ mag (Schlegel et al. 1998). Both HIZOA J1940+11 and
HIZOA J1914+10 appear isolated, in contrast to HIZOA J0630+08, and lie
in the outskirts of the Local Void.

\subsection{Survey Completeness}

Our ability to detect a galaxy in the survey depends on both its
integrated flux and velocity width. A galaxy with a high total flux
and a small to moderate velocity width can be identified most easily;
decreasing a galaxy's flux or increasing its velocity width would
generally make detection more difficult.  This, and previous
experience with HI surveys (e.g. Kilborn et al. 2002), leads us to
expect the survey to be approximately mean flux limited, where mean
flux is defined here to be the integrated flux divided by the 20\%
velocity width (or the 50\% velocity width in cases where the 20\%
width is not available). This expectation is confirmed by plotting the
number of survey galaxies with a given flux, $N(S)$, where $S$ is
defined to be the mean, integrated, or peak flux density.  We find
that $N(S)$ declines most steeply at low fluxes, and therefore best
defines the flux cutoff, when the flux is taken to be the mean flux
density.

A plot of $N(S)$ against $S$ is shown in Fig.5, with $S$ representing
the mean flux density as above. It is consistent with a Euclidean
power law, $N(S)\propto S_{\rm mean}^{-2.5}$, and a mean flux density
completeness limit of 0.022 Jy, assuming approximate spatial
homogeneity. The best-fit linear relationship between the mean flux
density and peak flux density of galaxies in the survey has a slope of
0.75.  The above completeness limit therefore corresponds roughly to a
peak flux density completeness limit of 0.029~Jy, which is $\sim
5\sigma$ above the rms noise of the survey, 6.0~mJy.

\begin{figure}
\epsscale{1.2}
\plotone{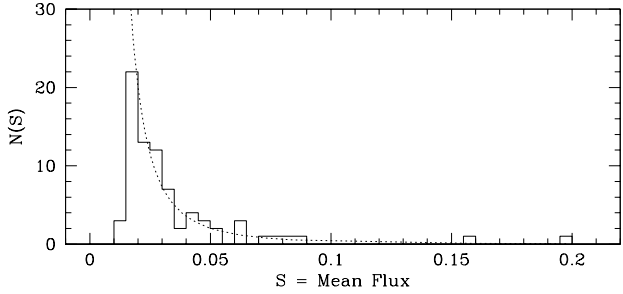}
\caption{Number of survey galaxies with a given mean flux density. 
A Euclidean power law, $N(S) \propto S_{\rm mean}^{-2.5}$, has been
fit to the data.  The approximate mean flux completeness limit,
0.022~Jy, was determined by averaging the fit of this power law to
regions of low, medium, and high rms noise.}
\end{figure}

In Figure 6, we plot as a solid histogram the area of the survey,
$A(\le S)$, at or below a given rms noise, $S$. The number of galaxies
detected at or below a given rms noise is also given as a dotted
histogram. We conservatively define the effective area of the survey
to be the area with an rms noise at or below 0.0073~Jy, 239~deg$^{2}$.
Inside this area, the limiting mean flux density of the survey,
0.022~Jy, corresponds to a 3$\sigma$, or higher, detection
(corresponding to 5$\sigma$, or higher, in peak flux density). This
area is less than the nominal survey area of
$4\times8\degr\times10\degr=320$ deg$^2$ partly because of edge
effects, and partly because of the increased continuum (particularly
residual continuum ripple) along the Galactic Plane.

To investigate the effect of the variable rms, we divide the survey
into three regions of equal area, but of different rms noise. We again
plot $N(S)$ against $S$ for each of the three regions and fit the
differential number counts with a Euclidean power law. We estimate
completeness limits of 0.019, 0.020 and 0.026~Jy for the regions of
low, medium, and high rms noise, respectively. As expected, the
completeness limit increases with rms noise. However, the variation is
sufficiently limited that, for the present purposes, a mean
completeness limit of 0.022~Jy adequately characterizes the survey.

Finally, it is worth noting that, in addition to good completeness
above a mean flux density of 0.022~Jy, this survey is expected to be
highly reliable. Galaxy identifications were made by three authors
independently, and the final galaxy list was made by a fourth author
in a conservative manner, guided by experience gained from follow-up
observations of surveys with similar noise characteristics (Henning et
al. 2000, Zwaan et al. 2004). We expect that very few, if any, of the
new galaxies presented here will be false detections, and therefore
the reliability of the catalog to be close to 100\%. Ultimately,
however, follow-up HI observations of some weaker objects will be
required to ensure that no false detections remain.

\begin{figure}[t]
\epsscale{1.0}
\plotone{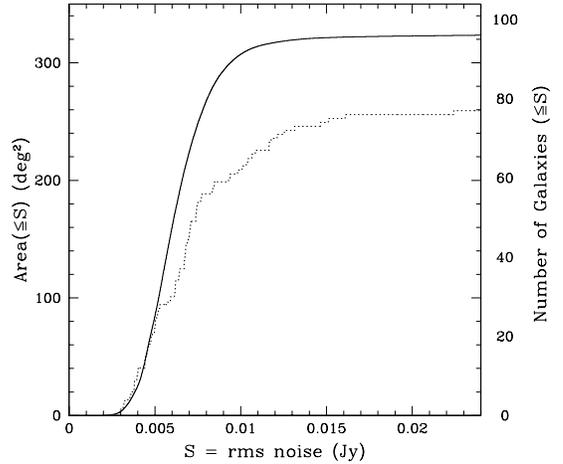}
\caption{Histogram of the survey area (solid line, left axis) and number 
of galaxies (dotted line, right axis) detected at or below a given rms
noise. The survey completeness is set to be the area of the survey
with an rms noise at or below 0.0073~Jy. }
\end{figure}

\begin{figure*}[t!]
\epsscale{0.8}
\plottwo{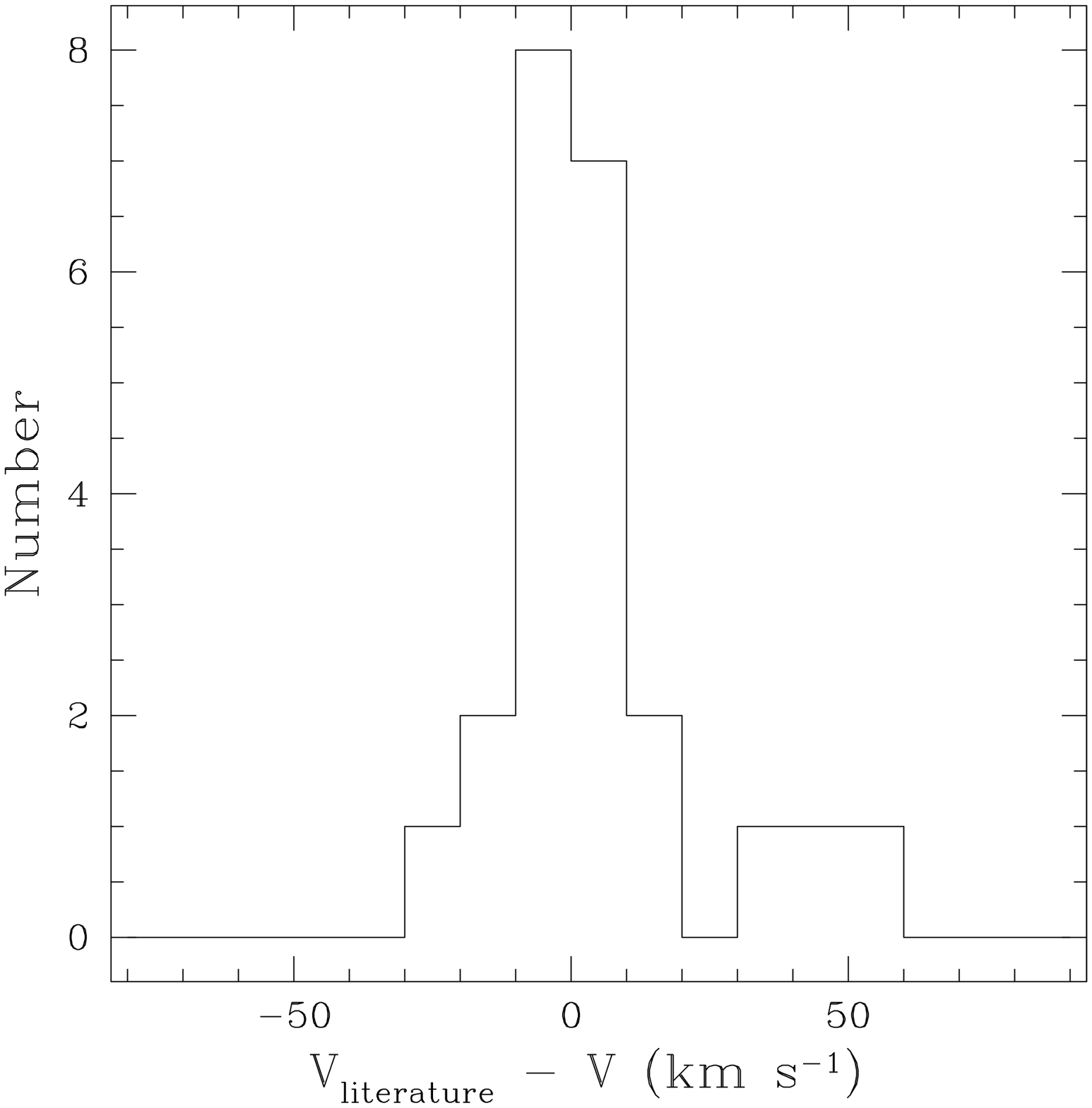}{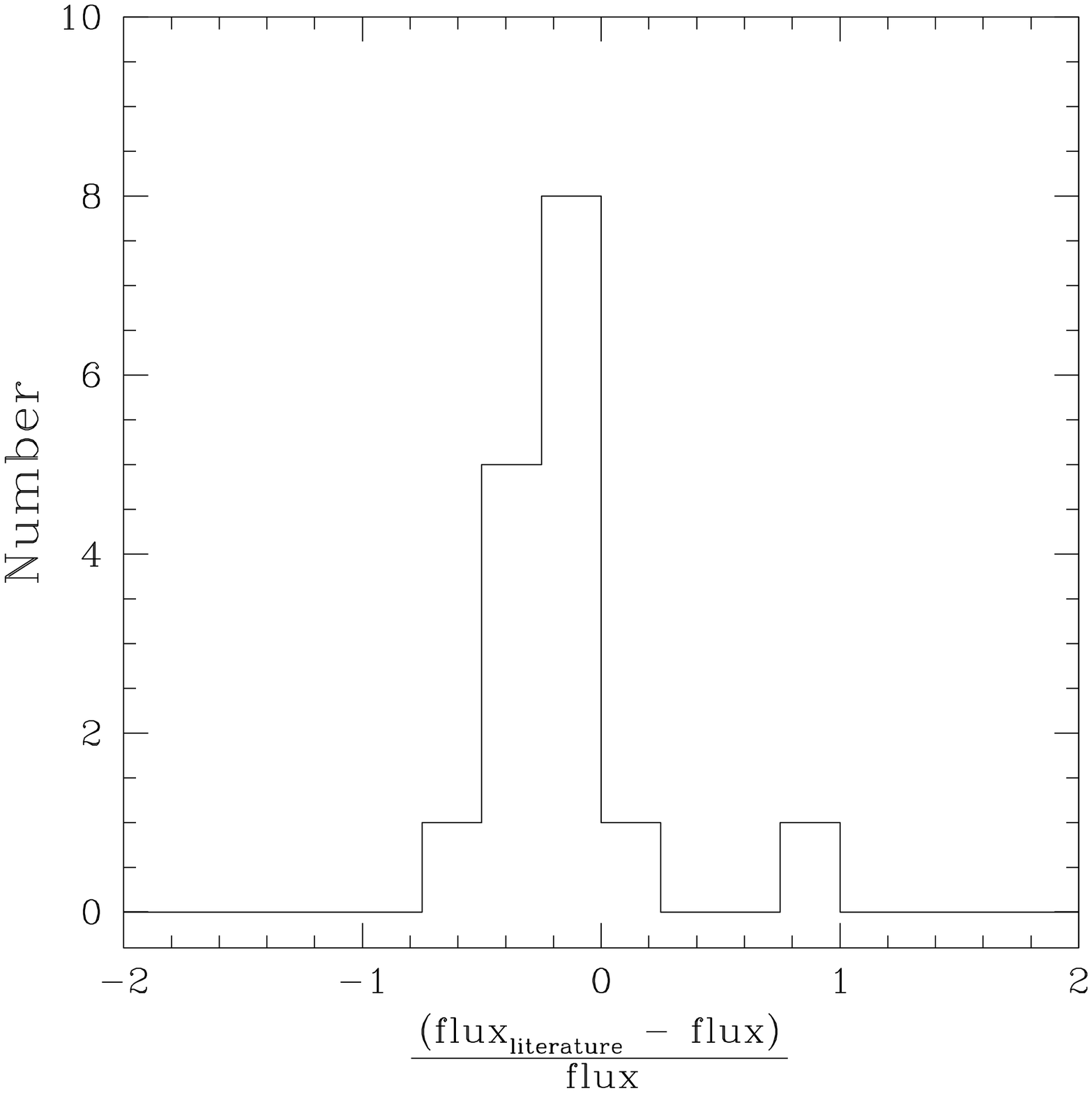}
\caption{Histograms of the offset between the HI velocities and fluxes measured
by our survey, and those from the literature. The mean offset in
velocity is $6$~km~s$^{-1}$ and that in flux is -16\%.}
\end{figure*}

\section{Multiwavelength Counterparts}

The region of sky covered by the northern ZOA survey overlaps slightly
with a number of other surveys.  These include four blind HI surveys,
the Arecibo Dual Beam Survey (ADBS; Rosenberg \& Schneider 2000), the
pilot NRAO 300-ft telescope survey (Henning 1992), the Dwingeloo
Obscured Galaxy Survey (DOGS; Henning et al. 1998, Rivers 2000), and
the HI Parkes Zone of Avoidance Shallow Survey (HIZSS; Henning et
al. 2000), although the overlap with the HIZSS is extremely small.
This region also coincides with a number of HI follow-up surveys and
several optical surveys, from which the ZOAG (see Lercher, Kerber, \&
Weinberger 1996), CGMW (see Roman, Iwata, \& Saito 2000), WEIN
(Weinberger 1980), and additional catalogs have been
compiled. Significant infrared coverage of these regions has been
provided by IRAS and 2MASS.

\subsection{Counterpart Search}

We searched for HI, optical, and infrared counterparts to the northern
ZOA galaxies using two methods. We first used the NASA/IPAC
Extragalactic Database (NED\footnote{This research has made use of the
NASA/IPAC Extragalactic Database (NED) which is operated by the Jet
Propulsion Laboratory, California Institute of Technology, under
contract with the National Aeronautics and Space Administration.}),
the 2MASS Extended Source Catalog (XSC\footnote{This publication makes
use of data products from the Two Micron All Sky Survey, which is a
joint project of the University of Massachusetts and the Infrared
Processing and Analysis Center/California Institute of Technology,
funded by the National Aeronautics and Space Administration and the
National Science Foundation.}), and the IRAS Point Source Catalog
(PSC) to conduct an automated search for the \textit{nearest}
potential counterparts to each HI galaxy. The fraction of these
counterparts likely to be associated with the HI galaxies was
estimated by a simulation, described in Appendix B.  Our second
approach was to determine the most likely counterparts by searching
for and visually inspecting all potential optical and infrared
counterparts, including (1) counterparts located further from the HI
position than those chosen by our previous search and (2) previously
unidentified galaxies in the Digitized Sky Survey (DSS\footnote{The
Digitized Sky Surveys were produced at the Space Telescope Science
Institute under U.S. Government grant NAG W-2166. The images of these
surveys are based on photographic data obtained using the Oschin
Schmidt Telescope on Palomar Mountain and the UK Schmidt
Telescope. The plates were processed into the present compressed
digital form with the permission of these institutions.})  POSS-II
near-infrared, red, and blue images.  Any galaxy visible in the DSS
bands was classified by us as an optical detection. The visual search
is, of course, subject to error; we have tried to minimize this error
by having multiple authors participate in the choice of
counterparts. Previous investigations of the offset between Parkes HI
sources and their counterparts (see Juraszek et al. 2000, Koribalski
et al. 2004, Ryan-Weber et al. 2002) indicate that while most
counterparts are found within $3^{\prime}$, offsets as high as
$5^{\prime}$ are not highly unusual. A conservative search radius of
10$^{\prime}$ was used for the initial searches to ensure that no
potential counterparts were missed.

\subsection{Previously Detected HI Galaxies}

Of the 77 galaxies detected here, 20 have been detected previously in
HI.  We have detected all galaxies from the ADBS, NRAO 300-ft, HIZSS,
and DOGS blind HI surveys that lie in our survey region, although
J0635+11 was resolved into two galaxies separated by $4^{\prime}$ by
the ADBS. We have also detected several galaxies from follow-up HI
surveys. Figure 7 illustrates the difference between our measured
systemic HI velocities and the HI velocities given by NED as well as
the fractional differences in flux between our survey and measurements
in the literature (Rosenberg \& Schneider 2000, Pantoja et al. 1997,
Pantoja et al. 1994, Theureau et al. 1998, Henning et al. 2000, Rivers
2000).  The distribution of velocity offsets has a mean of
$6$~km~s$^{-1}$ and a dispersion of 19~km~s$^{-1}$. The flux we
measure is offset by an average of -16\% from the values in the
literature, with a dispersion of 32\%.

The galaxies with the largest positive and negative flux offsets are
HIZOA J0654+08 and HIZOA J0622+11B, respectively.  Both were
previously observed as part of the Arecibo Dual Beam Survey.  The
small ADBS flux for HIZOA J0622+11B can potentially be explained by
the fact that ADBS flux measurements were underestimated if the source
was offset from the beam center (see Rosenberg and Schneider 2000).
We are confident in our fits to both galaxies, and are unable to
explain the large positive flux offset of HIZOA J0654+08.

Two of the remaining galaxies for which our flux is significantly
higher than that in the literature were identified by the Dwingeloo
Obscured Galaxies Survey. Both have strong clean profiles. The
discrepancies in flux most likely arise from the conservative way in
which the Dwingeloo galaxy fluxes were measured (Rivers (2000) defined
the total integrated flux to be the flux inside the $N_{HI} = 1.25
\times 10^{20}$~cm$^{-2}$ contour).  Also, as the Dwingeloo fluxes
were measured from VLA maps, any extended emission would have been
missed by the interferometer.

\subsection {Results of Counterpart Search}

We present in Table 2 the HI counterparts from other surveys of the
ZOA galaxies, along with the most likely optical, 2MASS, and IRAS
counterparts.  For the latter group, we include only those sources
that were visually verified to be likely counterparts. Potential
counterparts were discarded from this list if they did not appear to
be a galaxy, were offset from the HI position by a statistically
unlikely amount ($\ga 4^{\prime}$), or were of an unlikely galaxy
type. In seven cases, the 2MASS counterpart that was chosen by the
visual inspection was not the counterpart closest to the HI position.
All likely optical and IRAS counterparts (excluding the DSS galaxies
which were not found via the automated NED search) were the
counterparts nearest the HI position. In Table 3, we give the
positions of the newly detected DSS counterparts.

The spatial distribution of sources with multiwavelength counterparts
is shown in Figure 8, where solid circles represent galaxies with both
optical and infrared counterparts, solid triangles represent galaxies
with optical counterparts only, and solid stars represent galaxies
with infrared counterparts only. HI galaxies with no counterparts are
given as crossed open circles. We have overlaid contours of E(B-V) on
Figure 8 using the DIRBE/IRAS data.  Contour values of $1^{\rm
m}\!\!.0$ and $3^{\rm m}\!\!.0$ were used for the $l=36^{\circ}$ to
$52^{\circ}$ region, whereas values of $0^{\rm m}\!\!.3$ and $0^{\rm
m}\!\!.8$ were used for the contours of the $l=196^{\circ}$ to
$212^{\circ}$ region.

\begin{figure*}[t]
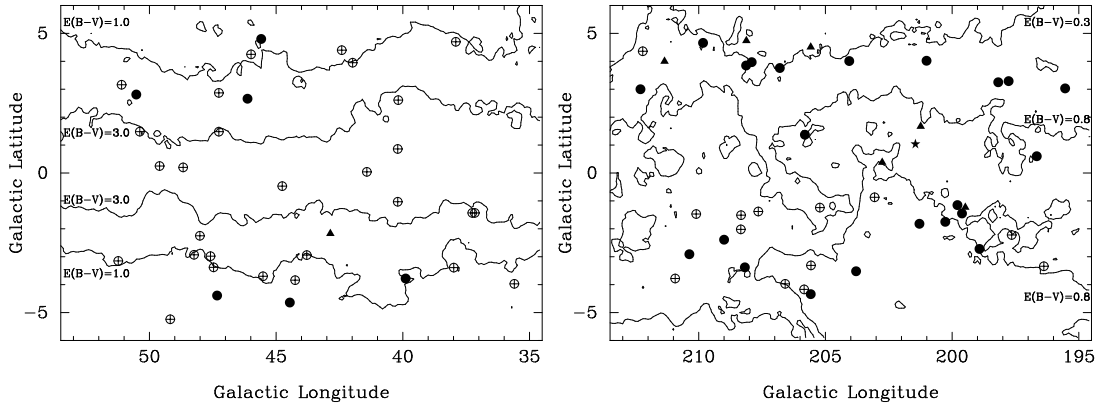

\epsscale{0.3}
\begin{center}
$\begin{array}{cc}
\includegraphics[angle=270,scale=0.30]{donley.fig8a.eps} &
\includegraphics[angle=270,scale=0.30]{donley.fig8b.eps}
\end{array}$
\caption{Spatial distribution of galaxies. HI galaxies with no
counterparts are represented by crossed open circles.  Solid circles
represent galaxies with both optical and infrared counterparts, solid
triangles represent galaxies with optical counterparts only, and solid
stars represent galaxies with infrared counterparts only. The
counterparts used in this plot are only those that remain after the
visual verification. The contours overlaid on the plot are contours of
E(B-V) (Schlegel et al. 1998), with values of $1^{\rm m}\!\!.0$ and
$3^{\rm m}\!\!.0$ for the $l=36^{\circ}$ to $52^{\circ}$ region, and
$0^{\rm m}\!\!.3$ and $0^{\rm m}\!\!.8$ for the $l=196^{\circ}$ to
$212^{\circ}$ region.}
\end{center}
\end{figure*}

In summary, 19 of our 77 galaxies have likely optical counterparts, 27
have likely 2MASS counterparts, and 11 have likely IRAS
counterparts. In addition, 16 galaxies have likely optical DSS
counterparts that are previously uncatalogued. The percentage of
counterparts found via the automated search that were still considered
likely after a visual verification is roughly consistent with the
expected percentages calculated by means of the simulations discussed
in Appendix B.


\section {Large-Scale Structure}

The spatial distribution of galaxies in our survey can be found in
Figures 9 through 12. In Figures 9 and 11, the HI galaxies detected in
our survey are given by solid points. We have plotted as open points
all galaxies in the Lyon-Meudon Extragalactic Database
(LEDA)\footnote{http://leda.univ-lyon1.fr -- the Lyon-Meudon
Extragalactic Database is supplied by the LEDA team at the Centre de
Recherche Astronomique de Lyon.} with known velocities. In each
velocity-slice, the galaxies with the lowest velocities are
represented by triangles; medium and high velocity galaxies are
plotted as squares and circles, respectively. The dotted box outlines
the region sampled by our survey.  In the cone diagrams of Figures 10
and 12, the HI galaxies are plotted as solid circles and LEDA galaxies
are plotted as open circles.  The survey region is again defined by
dotted lines.  The velocities in Figures 9, 10, 11, and 12 have been
converted to the Local Group standard of rest.  Void parameters are taken from
Fairall (1998) and are given in Table 4.

The $l=36^{\circ}$ to $52^{\circ}$ region lies near several voids,
including the Local, Microscopium, and Cygnus Voids.  At first glance
it appears that the galaxies detected in our survey may define the
edge of the Local and/or Microscopium Voids, but we caution that
preliminary results from the full ZOA survey indicate that the number
of galaxies rises slightly as one moves from our survey region towards
the Local and Microscopium Voids before dropping off at Galactic
longitudes of $10^{\circ}<l<20^{\circ}$.

\begin{deluxetable}{lcc}
\tablenum{3}
\tabletypesize{\small}
\tablewidth{0pt}
\tablecaption {New DSS Counterparts}

\tablehead{
\colhead{HIZOA}             &
\colhead{$\alpha_{2000}$}   &
\colhead{$\delta_{2000}$}    
}

\startdata
J0622+11A                    & 06 21 59.4  & +11 18 38  \\
J0634+09                     & 06 34 04.9  & +09 12 55  \\
J0635+11                     & 06 35 54.6  & +11 08 10  \\
J0641+01\tablenotemark{a}    & 06 41 04.4  & +01 50 24  \\
J0659+06                     & 06 59 38.1  & +06 27 19  \\
J0702+03                     & 07 02 50.4  & +03 11 14  \\
J1853+09\tablenotemark{a}    & 18 53 47.5  & +09 51 16  \\
J1917+07                     & 19 17 24.8  & +07 49 07  \\
J1927+09\tablenotemark{a}    & 19 27 49.5  & +09 27 32  \\
J1929+08                     & 19 29 19.9  & +08 02 42  \\
J1930+12\tablenotemark{a}    & 19 30 32.2  & +12 11 41  \\
\enddata

\tablenotetext{a}{We see a galaxy at this position, but cannot
determine whether it is the correct counterpart to the HI galaxy.}

\end{deluxetable}


\begin{deluxetable}{lrrrr}[t!]
\tablenum{4}
\tabletypesize{\small}
\tablewidth{0pt}
\tablecaption {Void Parameters from Fairall (1998)}

\tablehead{
\colhead{Name}                    &
\colhead{$l$}                     &
\colhead{$b$}                     &
\colhead{$cz$}                    &
\colhead{radius}                 \\
\colhead{}                        &
\colhead{(deg)}                   &
\colhead{(deg)}                   &
\colhead{(km~s$^{-1}$)}           &
\colhead{(km~s$^{-1}$)}       
}

\startdata
Aquarius Void             & 60   & -41  & 4500  & 1500        \\
Canis Major Void          & 229  & -13  & 5000  & 2500        \\
Capricornus Void          & 2    & -17  & 8500  & $\sim 2500$ \\
Cor Bor Void              & 58   &  37  & 5200  & 2000        \\
Cygnus Void               & 67   & -9   & 3500  & 1250        \\
Delphinus Void            & 59   & -6   & 2500  & 1500        \\
Gemini Void               & 172  & 9    & 3000  & $\sim 1250$ \\
Local Void                & 18   & 6    & 1500  & 1500        \\
Microscopium Void         & 10   & 1    & 4500  & 1750        \\
Orion Void                & 206  & -2   & 1500  & 750         \\
Taurus Void               & 167  & -29  & 4000  & $\sim 2000$ \\
\enddata
\end{deluxetable}


When visualized including even larger areas (not shown here), the
nearby detected galaxies that produce the distinct peak around
3000~km~s$^{-1}$ appear to be the smooth continuation of the sine-wave
like feature that can be traced across the whole southern sky
(Kraan-Korteweg, Koribalski, \& Juraszek 1999).  This wave feature
crosses the Milky Way in Puppis and continues towards the Antlia,
Hydra, and Centaurus structures before folding back across the
Galactic Plane in a structure called the Centaurus Wall by Fairall
(1998).  This wave encircles the Local Void while crossing the ZOA for
a third time towards the region inspected here.

The second peak in the velocity distribution around 4500~km~s$^{-1}$
also seems to be due to a clear filament present in both the middle
panel of Figure 9 as well as the redshift cones of Figure 10.  Here,
the detected galaxies seem to divide the Microscopium and
Cygnus/Delphinus Voids, the latter two of which are likely to form one
void rather than two distinct voids.

\begin{figure}
\epsscale{1.2}
\plotone{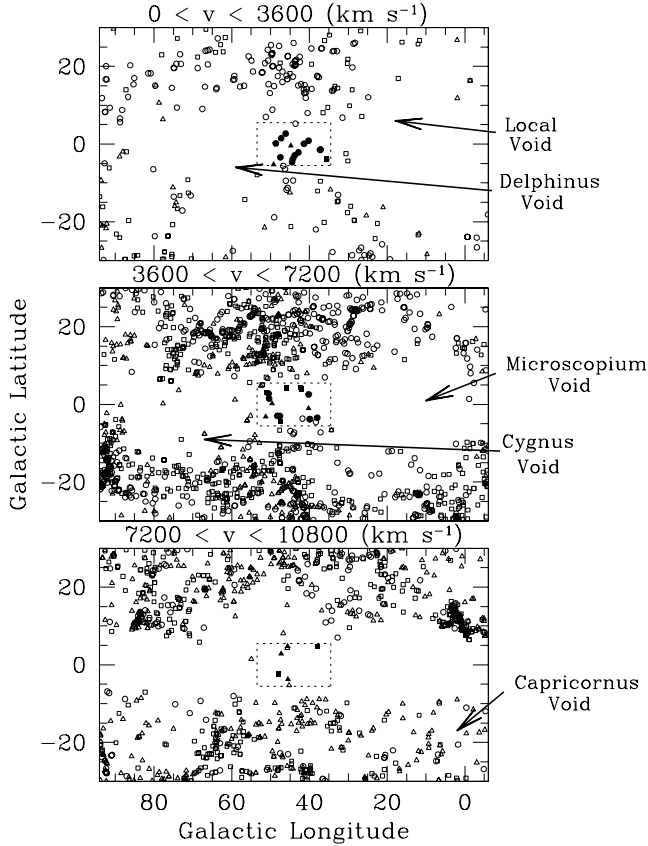}
\caption{Spatial distribution of galaxies in the $l=36^{\circ}$ to
$52^{\circ}$ region. The velocities are in the Local Group standard of rest. The
dotted line outlines the survey region. HI galaxies detected by this
survey are represented by solid points; LEDA galaxies are given by
open points. In each velocity range, triangles, squares, and circles
represent the galaxies in the low, middle, and high velocity bins,
respectively.  The dotted line outlines the survey region. }
\end{figure}

\begin{figure}
\epsscale{1.05}
\plotone{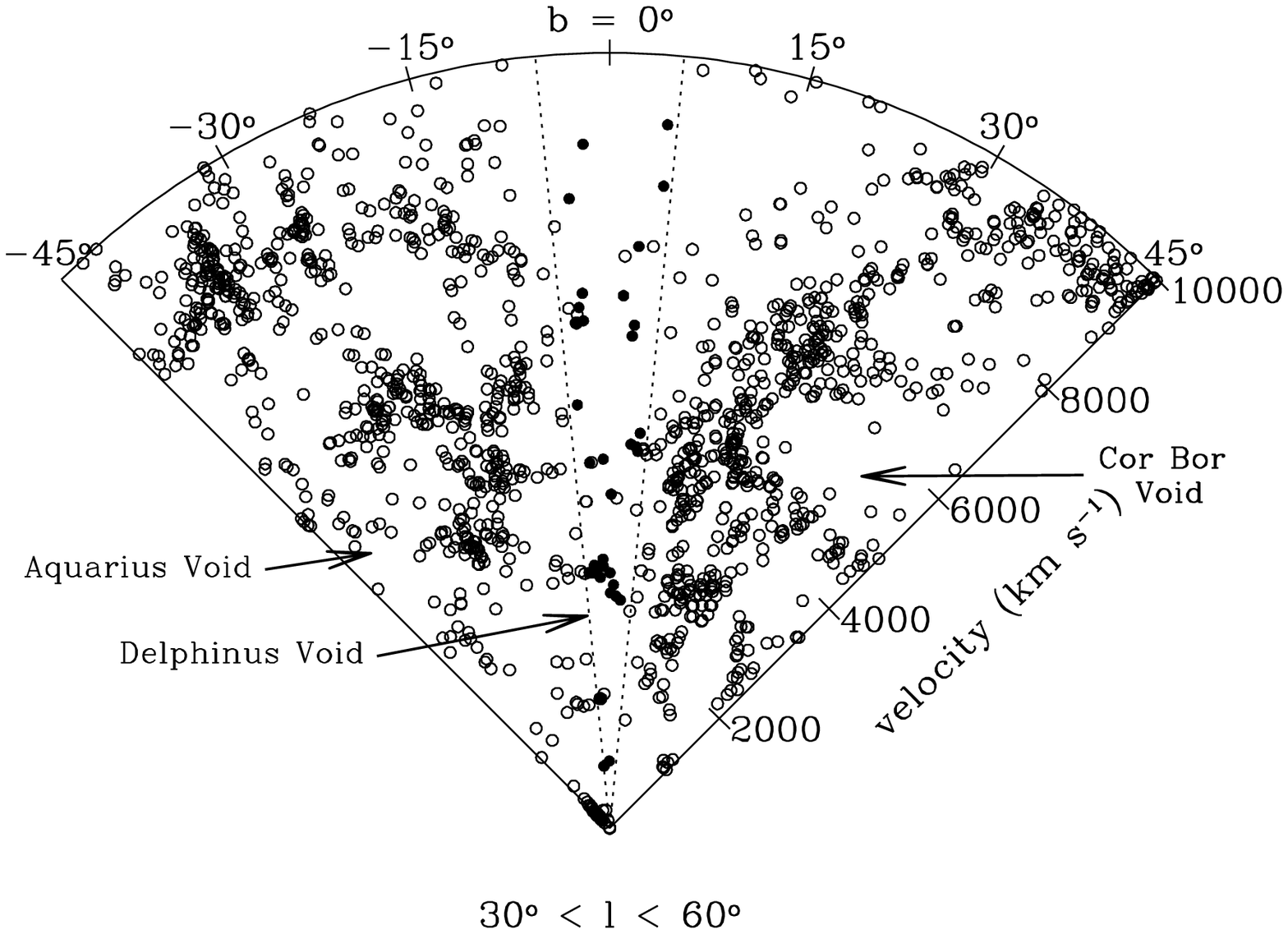}
\plotone{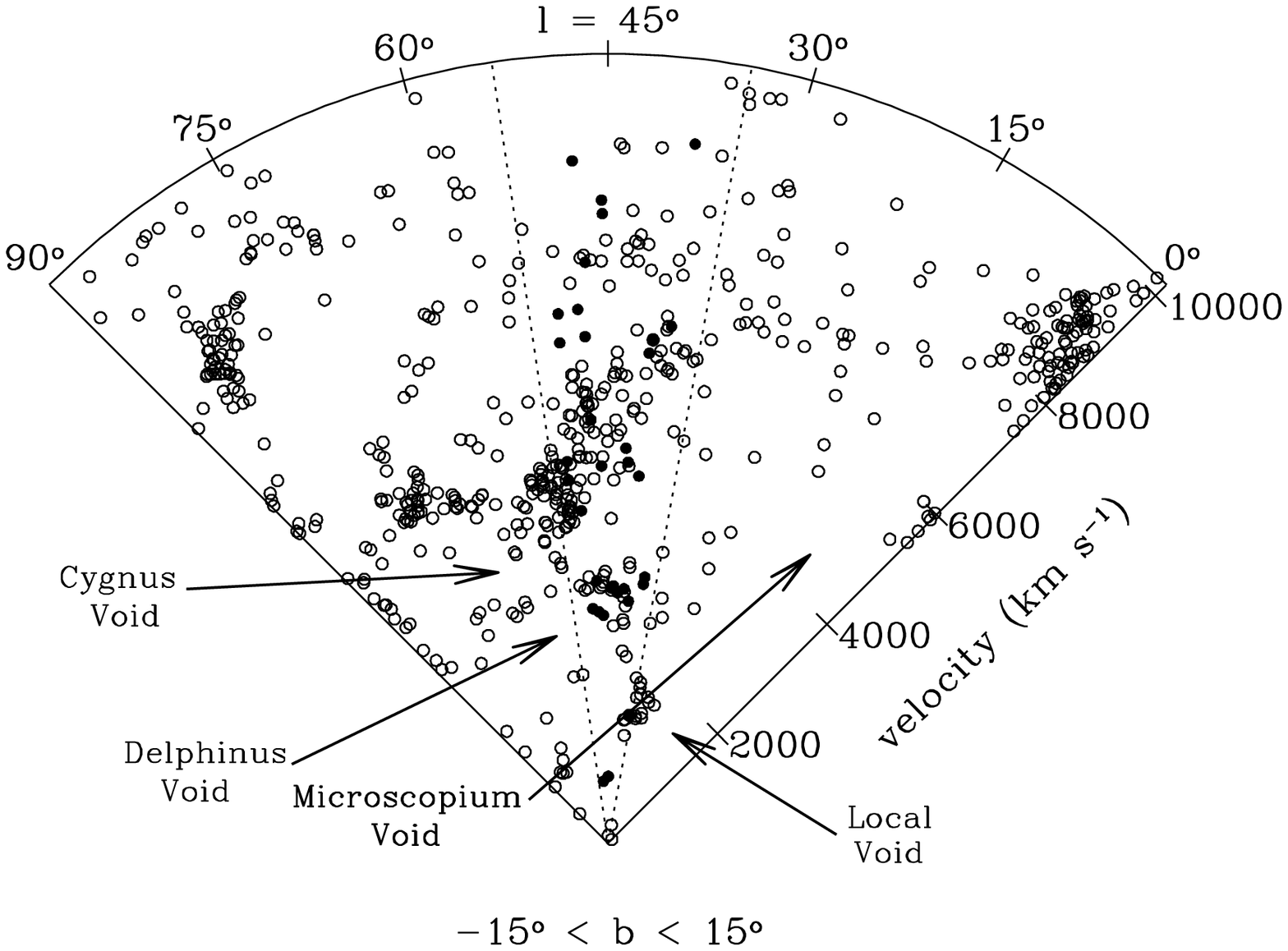}
\caption{Cone diagrams of galaxies in the $l=36^{\circ}$ to
$52^{\circ}$ region. The velocities are in the Local Group standard of rest. HI
galaxies detected by this survey are represented by solid circles;
LEDA galaxies are given by open circles. The survey region is defined
by dotted lines. }
\end{figure}

\begin{figure}
\epsscale{1.2}
\plotone{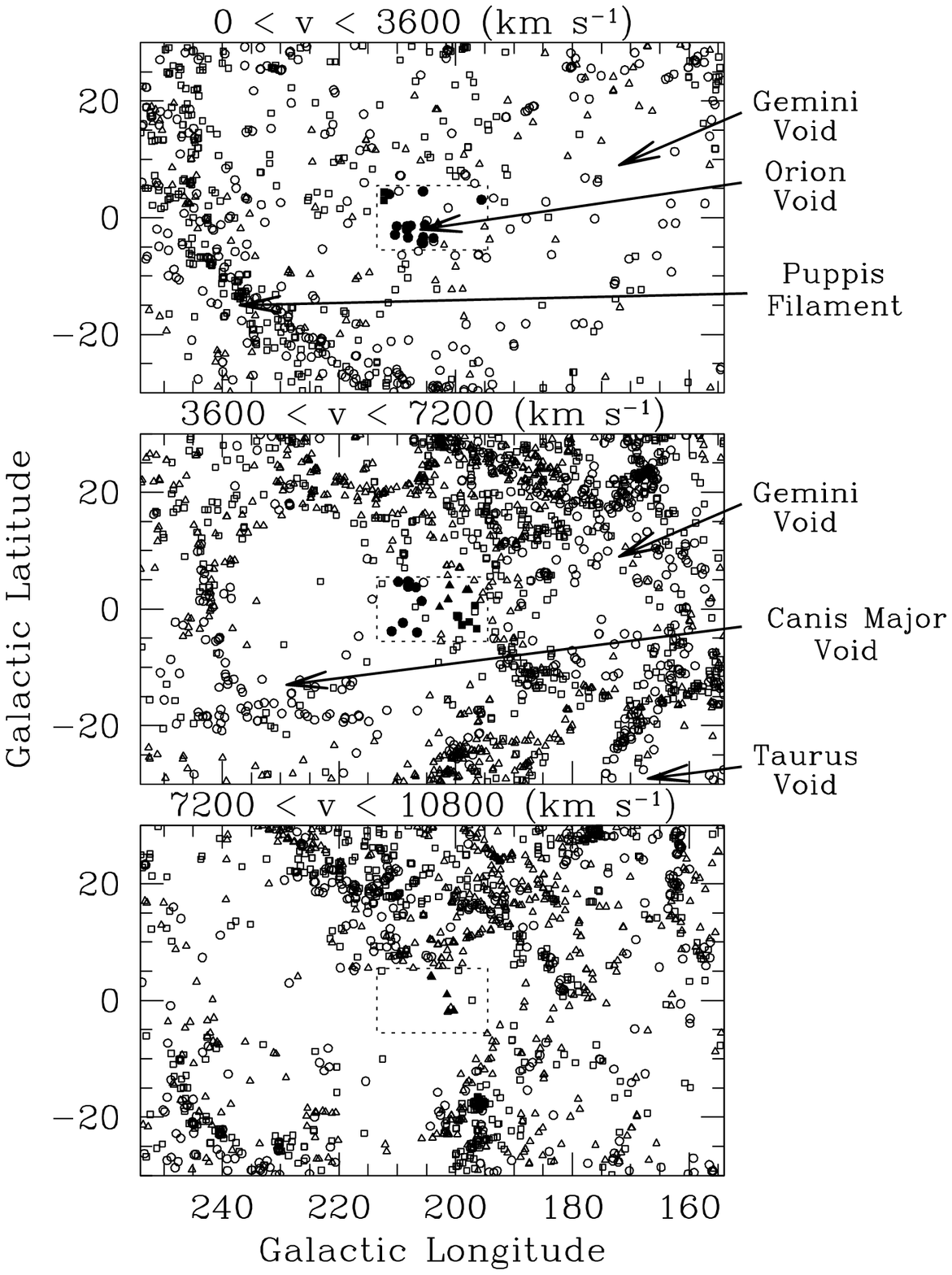}
\caption{Spatial distribution of galaxies in the $l=196^{\circ}$ to 
$212^{\circ}$ region. The velocities are in the Local Group standard of rest. The
dotted line outlines the survey region. HI galaxies detected by this
survey are represented by solid points; LEDA galaxies are given by
open points. In each velocity range, triangles, squares, and circles
represent the galaxies in the low, middle, and high velocity bins,
respectively.  The dotted line outlines the survey region.}
\end{figure}

\begin{figure}
\epsscale{1.05}
\plotone{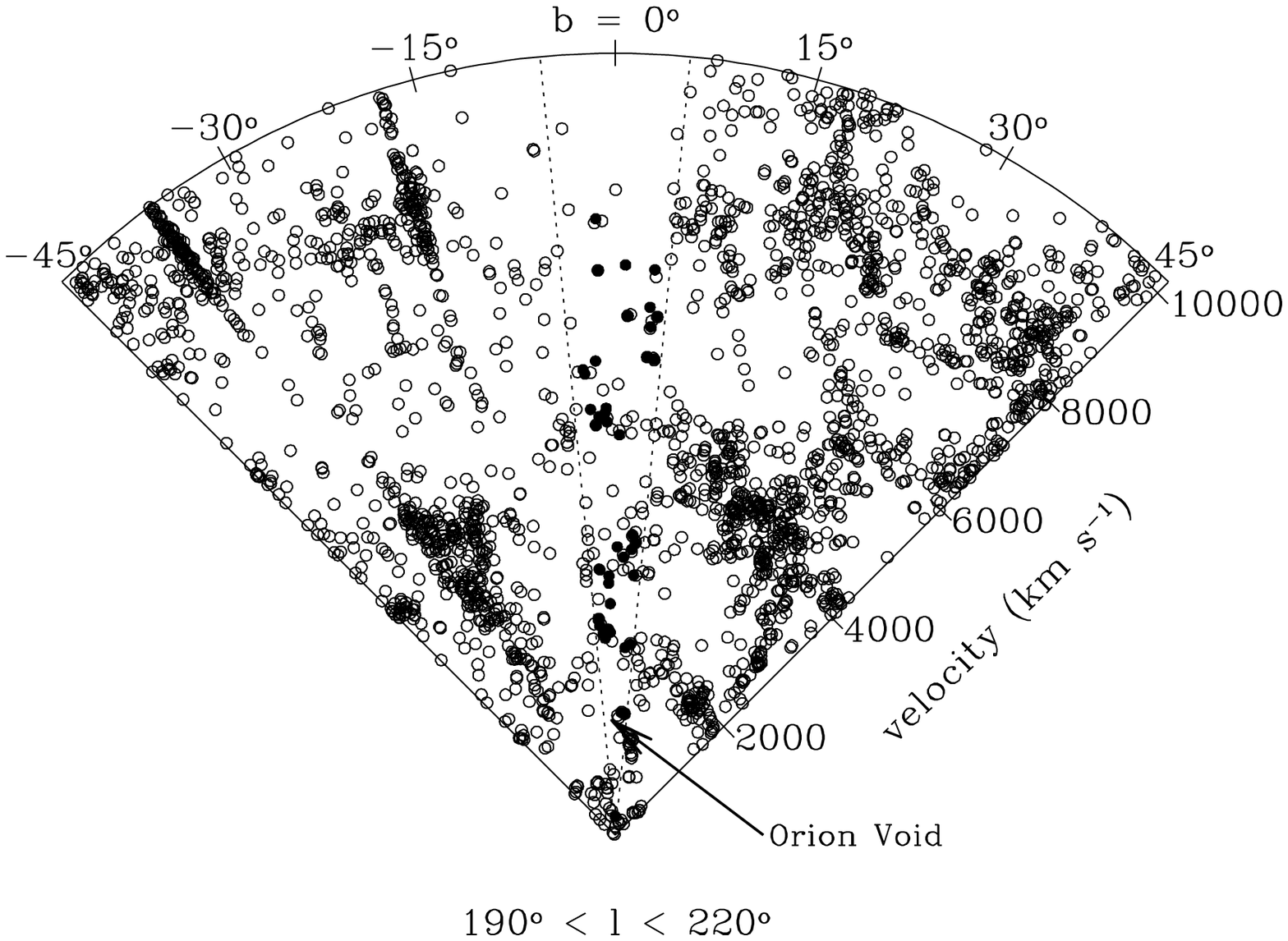}
\plotone{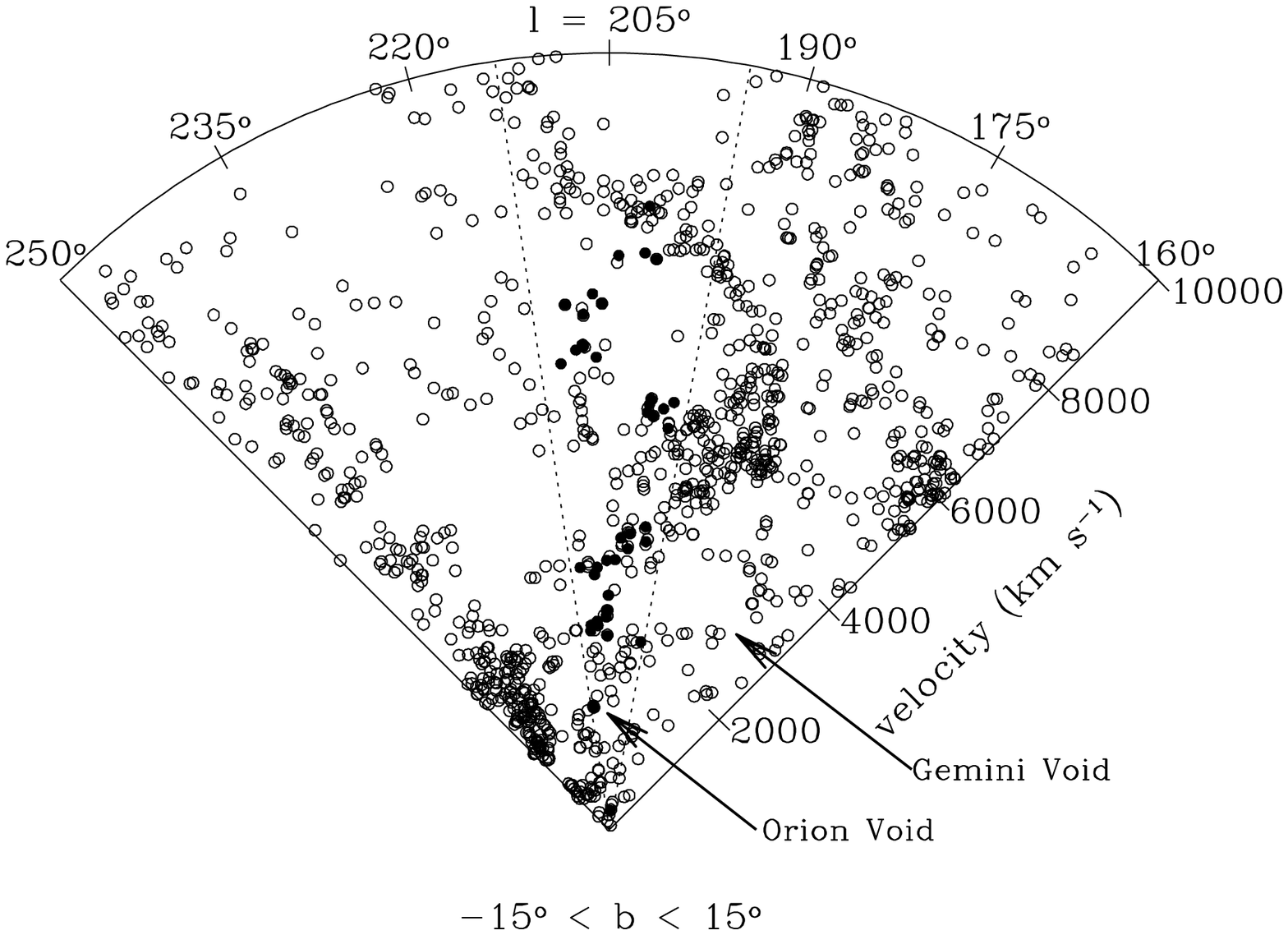}
\caption{Cone diagrams of galaxies in the $l=196^{\circ}$ to 
$212^{\circ}$ region. The velocities are in the Local Group standard of rest. HI
galaxies detected by this survey are represented by solid circles;
LEDA galaxies are given by open circles. The survey region is defined
by dotted lines.}
\end{figure}

There are several large-scale structures in the vicinity of the
$l=196^{\circ}$ to $212^{\circ}$ region, which lies between the
Perseus-Pisces supercluster and the Puppis filament. Two of the
nearest newly-detected galaxies, HIZOA J0700+01 and HIZOA J0705+02,
are found within 4.8 and 4.9~Mpc of the center of the Orion Void.
This void ($l, b, cz =206^{\circ}, -2.0^{\circ}, 1500$~km~s$^{-1}$)
was previously thought to have a radius of 750~km~s$^{-1}$, or 10~Mpc
(Fairall 1998). The LEDA Database, however, contains 42 galaxies that
lie between 5~Mpc and 10~Mpc of the void center, and two that lie at
smaller radii (4.6 and 4.9~Mpc).  It therefore appears that the true
radius of this void is approximately 5.0~Mpc and that the two galaxies
detected here lie along its edge.

The galaxies between 2500 and 5500~km~s$^{-1}$ seem to provide the
border of two voids, the Gemini Void, now quite prominent in the upper
panel of Figure 11, and the Canis Major Void, which now seems slightly
smaller and centered at higher latitudes ($l, b, cz =220^{\circ},
0^{\circ}, 5000$~km~s$^{-1}$) as compared to its location as given by
Fairall 1998 ($l, b, cz =229^{\circ}, -13^{\circ}, 5000$~km~s$^{-1}$).
The distinct gap between 4000 and 5000~km~s$^{-1}$ in the velocity
distribution (Fig. 4) is due to the fact that the continuous part of
the border of the Canis Major Void between 4000 and 5000~km~s$^{-1}$
falls just outside of our survey regions and crosses the ZOA at quite
constant velocities.


\section{Mass Function}

\begin{figure*}
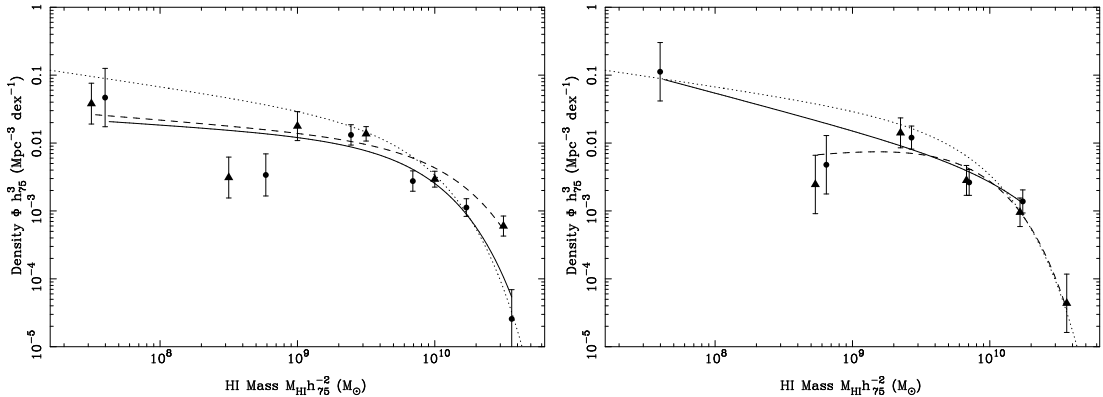

\begin{center}
$\begin{array}{cc}
\includegraphics[angle=270,scale=0.30]{donley.fig13a.eps} &
\includegraphics[angle=270,scale=0.30]{donley.fig13b.eps}
\end{array}$
\caption{(a) HI mass function of the survey galaxies. The circles represent 
the mass function calculated from the standard $1/V_{\rm max}$
method. The solid line is a least-squares fit of a Schechter function
for this method. The modified $1/V_{\rm max}$ method is represented by
triangular points. The Schechter function fit for this method is given
by a dashed line. (b) HI mass function of the two survey regions.  The
mass function of the $l=36^{\circ}$ to $52^{\circ}$ region is
represented by circular points and a solid line. The mass function of
the $l=196^{\circ}$ to $212^{\circ}$ region is represented by
triangular points and a dashed line. The bin size is 0.5 dex; bins are
plotted if one or more galaxies are present. Points are plotted at the
mean mass within the bin for the standard method, and at the mean bin
position for the modified method. For both plots, the dotted line
represents the mass function from the HIPASS BGC (Zwaan et al. 2003).}
\end{center}
\end{figure*}

\begin{deluxetable*}{llllll}
\tablenum{5}
\tabletypesize{\small}
\tablewidth{0pt}
\tablecaption {Schechter Parameters}

\tablehead{
\colhead{Method}                            &
\colhead{log $\Theta^{\ast}$}               &
\colhead{$\alpha$}                          &
\colhead{log $M_{\rm HI}^{\ast}$}           &
\colhead{log HI density\tablenotemark{a}}   &
\colhead{log HI density\tablenotemark{b}}                 
}

\startdata
$1/V_{\rm max}$           & $-2.33 \pm 0.33$  & $1.13 \pm 0.36$  & $9.85 \pm 0.19$   & $7.56 \pm 0.42$  & $7.55^{+0.07}_{-0.08}$    \\
Modified $1/V_{\rm max}$  & $-2.35 \pm 0.32$  & $1.16 \pm 0.27$  & $10.05 \pm 0.20$  & $7.75 \pm 0.37$  & \nodata                   \\
\enddata

\tablenotetext{a}{HI mass density calculated from Schechter parameters (log $\msun$ Mpc$^{-3}$)}
\tablenotetext{b}{HI mass density calculated from $\Sigma {\rm{M_{\rm HI}}\over v_{\rm max}}$ (log $\msun$ Mpc$^{-3}$)}

\end{deluxetable*}

Over the past several years, a number of attempts have been made to
better define the HI mass function.  Most recently, the HIPASS Bright
Galaxy Catalog (BGC, Zwaan et al. 2003) and the ADBS (Rosenberg \&
Schneider 2002) have been utilized. Prior to these surveys, which
cataloged 1000 and 265 galaxies, respectively, the HIPASS SCC (Kilborn
2000), the HIPASS HIZSS (Henning et al. 2000), and the AHISS (Zwaan et
al. 1997) provided the best estimates to the mass function. The
sensitivity and coverage of our survey do not allow us to estimate the
mass function any more precisely than these recent measurements,
especially at the low-mass end where the slope of the mass function is
most uncertain. Determining the mass function is still of interest,
however, as it allows an investigation of the comparative mass density
and distribution of galaxies in our survey region, as well as a rough
comparison with previous estimations.

We calculate the HI mass function in two ways, the standard $1/V_{\rm
max}$ method (Schmidt 1968) and a refined $1/V_{\rm max}$ method
(Saunders et al. 1990). The latter incorporates the density profile of
the galaxies into the calculation to reduce the effect of clustering
and large-scale structure. The standard $1/V_{\rm max}$ method
non-parametrically determines the mass function by weighting each
galaxy's mass by the inverse of the maximum volume in which it could
be detected by the survey. This method, however, is potentially
affected by large-scale structure, as it assumes that the galaxy
population has a homogeneous distribution. The Saunders et al. method
is able to remove the majority of the effects of large-scale structure
by first using a maximum likelihood technique to determine the radial
density distribution of the sample, $p(r)$.  This density profile is
used to define an effective volume, $V_{\rm eff} = \int_{0}^{\rm
r_{\rm max,i}} r^{2}p(r)dr$, which replaces $V_{\rm max}$ in the
$1/V_{\rm max}$ method. In both cases, the resulting mass function can
be fit by a Schechter (1976) function:

$$\theta(M_{\rm HI})dM_{\rm HI} = \theta^{\ast} ({{M_{\rm
HI}}\over{M_{\rm HI}^{\ast}}})^\alpha {\rm exp} (-{{M_{\rm
HI}}\over{M_{\rm HI}^{\ast}}}) dM_{\rm HI}$$

The HI mass function can be found in Figure 13, where the data points
from the standard $1/V_{\rm max}$ method are given by circles and the
fit by a solid line.  The modified $1/V_{\rm max}$ method is
represented by triangular points and a dashed line. For comparison,
the HI mass function determined from the BGC has been included on the
plot, as a dotted line. Only those galaxies with a mean flux above our
completeness limit of 0.022~Jy were included in the mass function
calculation.  A bin size of 0.5~dex was used for both techniques. For
the standard $1/V_{\rm max}$ method, the points are plotted at the
mean mass of the galaxies in each bin. Missing points indicate that we
detect no galaxies in that bin.  For the modified $1/V_{\rm max}$, the
points represent the middle of the mass bin. The galaxy J0635+11,
which was resolved into two galaxies by the ADBS, does not affect our
mass function, as its mean mass falls below our cutoff. For a
discussion on the effects of confusion on HI mass functions, see Zwaan
et al. (2003).

The Schechter parameters from the two methods as well as the derived
HI mass densities can be found in Table 5. The two methods give
similar mass functions, suggesting that large-scale structure in these
regions does not strongly affect the calculation of the mass function.
The largest deviation between the two methods occurs at the high-mass
end, where the standard method is in better agreement with the BGC
mass function than the modified method. In addition, the standard
$1/V_{\rm max}$ method gives a $<{V\over{V_{\rm max}}}>$ of $0.54 \pm
0.04$, indicating that spatial inhomogeneity will not affect the
results greatly. Both methods of estimating the mass function give a
function that falls below that of the BGC at the low mass end, with
the largest discrepancy coming from an underdensity of galaxies of
mass $M\sim 10^{8}-10^{9}$ \msun.  This offset may be due to an
underdensity of mass in the regions examined.  Zwaan et al. (2003)
have shown that the four quadrants of the BGC give mass functions with
similar values of $M^{\ast}$ but with varying slopes and
normalizations with a degree of variation similar to what we observe
between our mass function and that derived from the BGC.



We also include in Figure 13 a plot of the mass functions for the two
separate survey regions. The $l=36^{\circ}$ to $52^{\circ}$ region is
given by circular points and a solid line.  The $l=196^{\circ}$ to
$212^{\circ}$ region is given by triangles and a dashed line. Due to
the low number of galaxies, these mass functions are not highly
reliable, but can still be used for general comparison purposes. The
mass function for the $l=196^{\circ}$ to $212^{\circ}$ region fits the
BGC mass function remarkably well in the high-mass regime, whereas the
$l=36^{\circ}$ to $52^{\circ}$ region mass function provides the best
fit to the BGC mass function in the low mass regime.


\section{Summary}

We have detected 77 HI galaxies in the northern extension of the
Parkes ZOA survey. The survey has a median rms noise of
6.0~mJy~beam$^{-1}$, is approximately complete to a mean flux density
of 22~mJy, and has an effective area of 239~deg$^{2}$. Of the 77
galaxies detected here, 19, 27, and 11 have likely optical, 2MASS, and
IRAS counterparts already cataloged, respectively.  In addition, a
further 16 have likely optical counterparts visible in the Digitized
Sky Survey.  Twenty have been previously detected in HI. The spatial
and velocity distributions of the galaxies in the $l=36^{\circ}$ to
$52^{\circ}$ region reveal several filaments that cross the Galactic
plane, the nearest of which seems to be the continuation of a
sine-wave like feature that can be traced across the whole southern
sky. The medium-velocity galaxies in the $l=196^{\circ}$ to
$212^{\circ}$ region seem to provide the border of the Gemini and
Canis Major Voids, whereas the detection of two new low-velocity
galaxies within the Orion Void help to redefine its radius. Of
particular interest is the galaxy HIZOA J0630+08 ($l,b$ =
203$^{\circ}$, --0.9$^{\circ}$), which has a velocity of $367\pm
1$~km~s$^{-1}$ in the heliocentric standard of rest.  We suggest that
it belongs to the nearby Orion Group which includes a small number of
dwarf galaxies. An HI mass function was derived using both the
standard $1/V_{\rm max}$ method and a revised $1/V_{\rm max}$ method
that is less sensitive to large-scale structure. Both methods produce
similar mass functions whose normalizations fall slightly below that
of the HI Parkes Bright Galaxy Catalog. The observations for the
southern Parkes ZOA survey are complete and analysis is currently
underway. Upon completion, this survey will be combined with the
northern extension studied here, which should considerably increase
our understanding of the structure and dynamics of the obscured Local
Universe.


\acknowledgments 
We thank the staff of the Parkes radio telescope. Also Will Saunders
for his implementation of the maximum likelihood $1/V_{\rm max}$
method. We gratefully acknowledge financial support from the
Australian-American Fulbright Commission (JLD) and CONACYT research
grant 40094-F (RCKK and JMII).

\begin{appendix}
\section{A. Errors in HI Profile Parameters}

To estimate the errors on the HI parameters, each galaxy profile was
smoothed using a Savitzky-Golay smoothing filter with a smoothing
width of 8 velocity channels (Press et al. 1992, Sec. 14.8). Fifty
simulated spectra were then created for each galaxy by adding random
Poisson noise to the smoothed galaxy spectrum; the rms of this random
noise was set equal to that of the original galaxy spectrum. The
errors on the HI parameters were then taken to be the median absolute
offsets between the parameters of the 50 simulated spectra and those
of the smoothed galaxy spectrum.

To better understand the errors on the HI parameters and the error
simulation technique, we investigate below three additional aspects of
the error estimation: (1) the effect of S/N on the measured errors,
(2) the accuracy of the smoothing filter in preserving the galaxy
parameters and (3) the biases introduced to the errors by the
simulation process.  We explore these issues using the spectrum of
J1853+09, which has a S/N of 24. This spectrum, and the smoothed
profile created using the Savitzky-Golay smoothing filter, are shown
in Figure A1. A spectrum with a high S/N was chosen to ensure that the
smoothed spectrum, used below as the template, was as representative
as possible of a real galaxy profile.

To explore the effect of S/N on the errors, we began by running the
error simulation, as described in \S 3.1, on the template. The
simulation was run 21 times; on each occasion, the rms of the random
Poisson noise was adjusted to give a desired S/N. The resulting set of
21 simulations have S/N ratios ranging incrementally from 3 to 24. The
errors on the HI parameters, as a function of S/N, are given in Figure
A2(a), where the points represent the median errors and the error bars
represent the median absolute value of the errors (in both the +y and
-y directions). The errors begin to rise at a S/N of 10, and do so
quite substantially for S/N $< 5$. This result is in agreement with
that of Roth, Mould, \& Staveley-Smith (1994), who performed a similar
set of simulations and found that the errors on HI line widths are
particularly unreliable for S/N $< 5$. Of the 77 galaxies in our
survey, none have a peak S/N $< 5$, 21 have $5 <$ S/N $< 10$, and 56
have S/N $> 10$.

Each of the 21 error simulations described above created 50 simulated
galaxy profiles with a given S/N. To investigate the ability of the
smoothing filter to preserve the galaxy profile, we next smoothed each
set of 50 simulated profiles using the Savitzky-Golay smoothing
filter. The HI parameters of the smoothed spectra were then measured
by {\sc mbspect}. Figure A2(b) illustrates the difference between the
HI parameters of the spectra for which noise was added and then
removed and the HI parameters of the original smoothed profile. There
are several systematic offsets. For S/N $\ge 6$, the peak flux density
from the simulated profiles is lower than the peak flux density of the
original smoothed profile, as expected from the smoothing
process. This offset is quite low, however, tending towards 5\% at
high S/N. The total flux, systemic velocity, 50\% velocity width, and
20\% velocity widths are all slightly larger in the simulated smoothed
profiles, but remain relatively constant down to S/N ratios of 5,
below which the differences in peak flux density, total flux, and 20\%
velocity width increase.  While it is not surprising that the
smoothing filter increases the total flux and velocity widths, the
increase in the systemic velocity is unexpected.  This change,
however, is due to the particular spectrum being used here, and will
likely vary with different profiles.  As can be seen in Figure A1, the
overestimation of the velocity at the high-velocity end of the profile
is larger than the underestimation at the low-velocity end, leading to
a slight ($\sim 2$~km~s$^{-1}$) overestimation of the systemic
velocity.

We lastly investigate how the changes in smoothed parameters discussed
above affect the error estimation. To do so, the error simulations
were run on each of the 50 simulated galaxy profiles with a given S/N
ratio. Recall that these 50 simulated spectra were created by adding
random noise with a given rms to the template. By running the
simulations on these 50 profiles, we remove the noise that we added
(the step explored in the last paragraph) and then re-add it to the
smoothed spectra. If no biases are introduced by the simulation
process, the errors now measured should be equal to the errors
estimated by running the simulations on the original spectrum of
J1853+09.  Figure A2(c) illustrates the differences between the errors
derived from the simulated profiles and those estimated from the
template. The largest bias introduced by the simulation process is the
underestimation of the error on the 20\% velocity width for spectra
with low S/N ratios. At S/N $\ge 10$, the error on the peak flux is
overestimated by $\sim 1.6$\% and the error on the total flux is
underestimated by $\sim 0.8$\%.  The errors on the systemic velocity,
50\% velocity width, and 20\% velocity width are offset by an average
of -0.2~km~s$^{-1}$, -0.9~km~s$^{-1}$, and 0.2~km~s$^{-1}$.  At
moderate to high S/N, these offsets are all well within the median
errors measured for the galaxies in this sample (see
\S3.1), confirming the reliability of this process in estimating
errors.

\begin{figure}
\figurenum{A1}
\epsscale{0.35}
\plotone{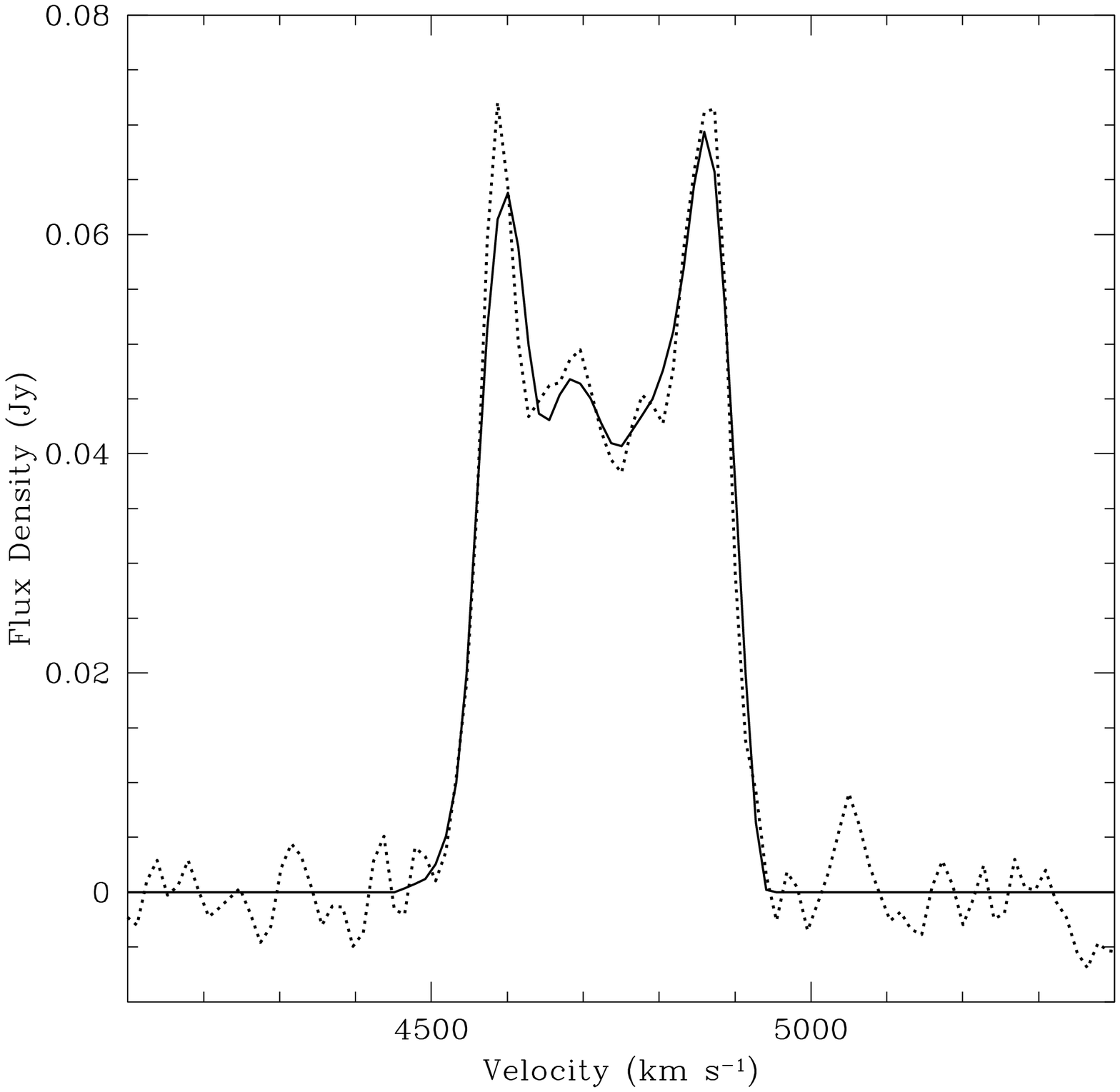}
\caption{Spectrum (dotted line) and Savitzky-Golay smoothed profile 
(solid line) of J1853+09.}
\end{figure}

\begin{figure}
\figurenum{A2}
\epsscale{0.5}
$\begin{array}{ccc}
\includegraphics[angle=0,scale=0.70]{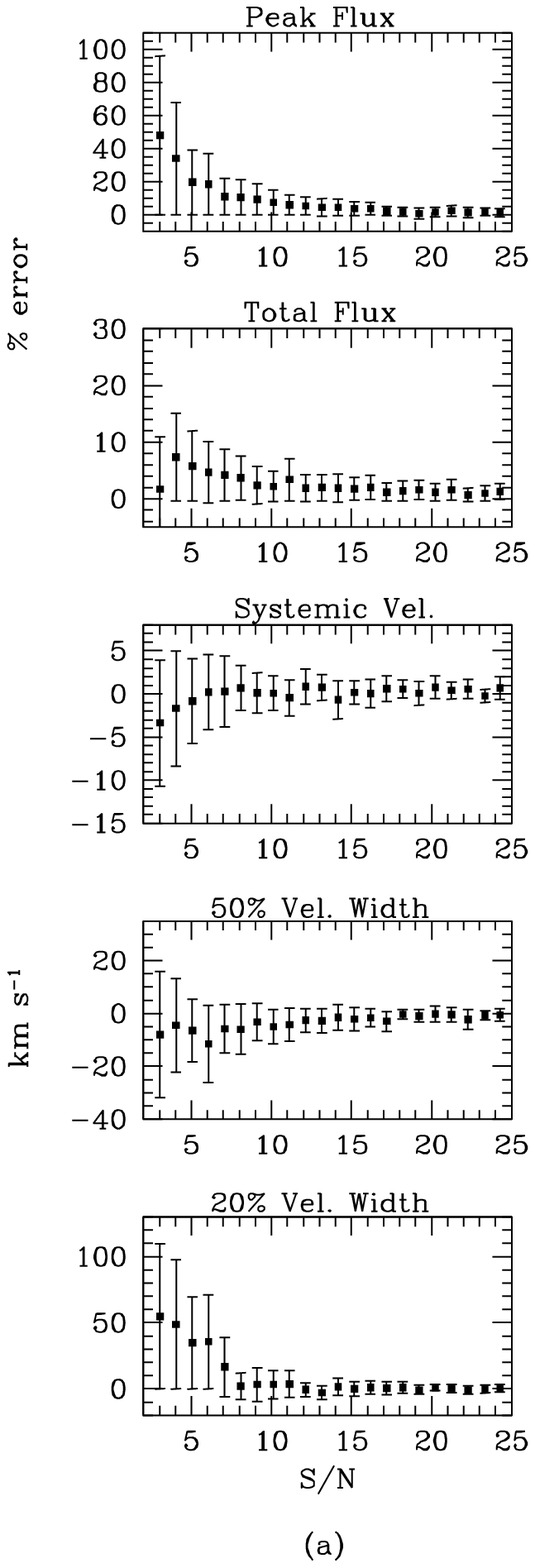} &
\includegraphics[angle=0,scale=0.70]{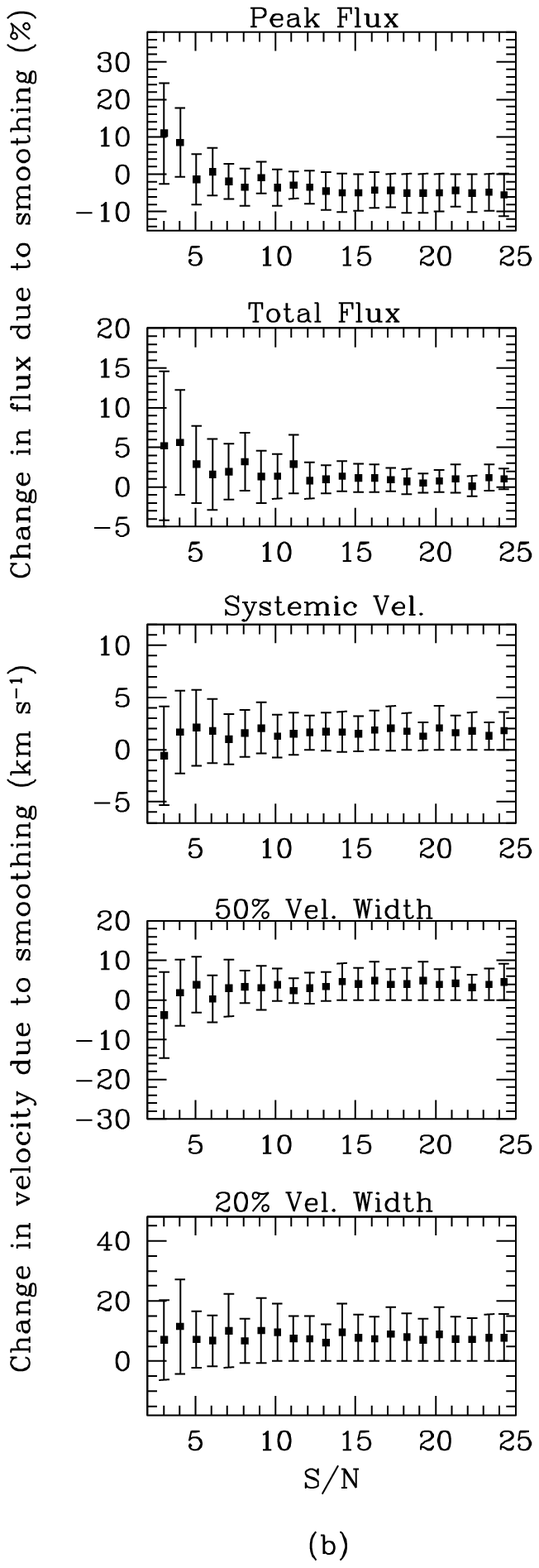} &
\includegraphics[angle=0,scale=0.70]{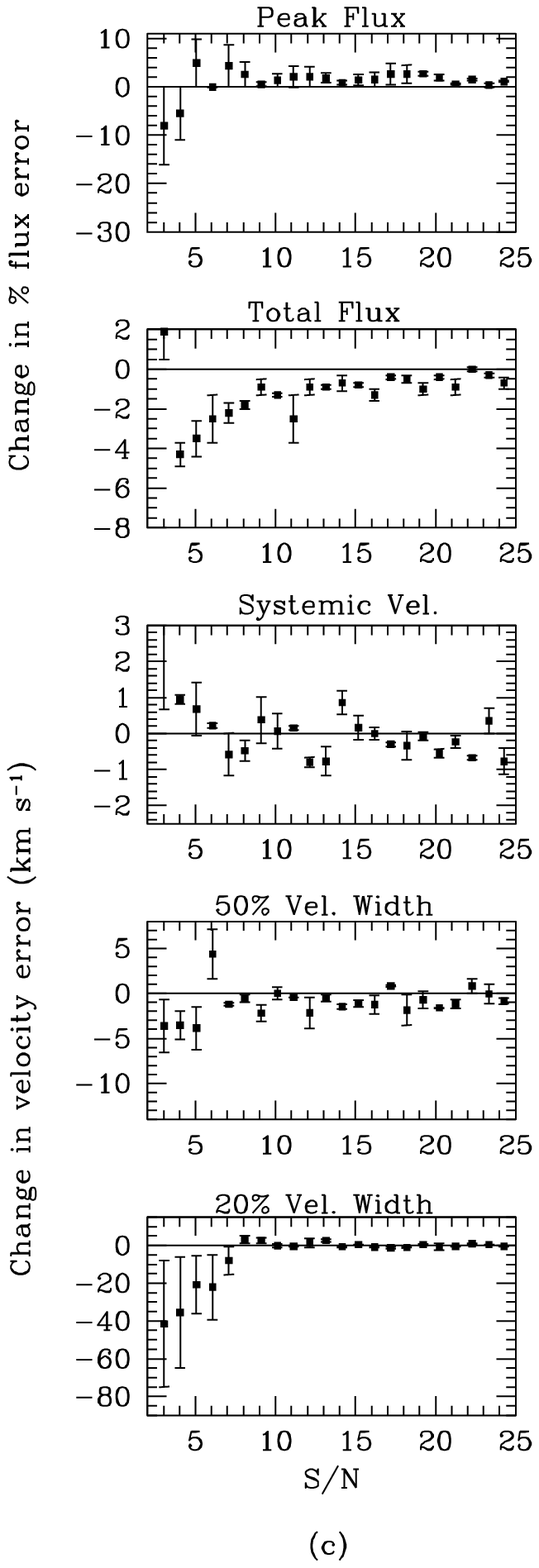}
\end{array}$
\caption{(a) The errors on measured HI parameters for the template as 
a function of S/N; (b) the difference between the parameters of the
smoothed simulated profiles and the original parameters of the
template; (c) the difference between the absolute value of the HI
errors measured from the simulated spectra and those measured from the
template. The points represent the median errors and the error bars
(in both the +y and -y directions) represent the median absolute value
of the errors.}
\end{figure}

\section{B. Counterpart Likelihood Simulations}

To investigate the likelihood that the \textit{nearest} potential
optical and infrared counterparts within both $3^{\prime}$ and
$5^{\prime}$ of the HI position are real counterparts of the HI
galaxies, a simulation was performed. In this simulation, fiducial HI
detections were placed across our survey region. The positions of
these simulated detections were defined by a grid of 120 positions at
20 arcminutes intervals surrounding each of the 77 galaxies.  Using
NED, the 2MASS XSC, and the IRAS PSC, counterparts of these false
objects were searched for. By measuring the number of objects near
these random positions, we determined the number of spurious
counterparts expected for our real galaxies, and thus the probability,
$P$, that the potential counterparts we detect are in fact related to
the HI galaxies.
\begin{equation}
P = {{N_{\rm counterparts} - N_{\rm unrelated}}\over{{N_{\rm
counterparts}}}}
\end{equation}
\begin{equation}
N_{\rm unrelated} = {N_{\rm simulated\ galaxies\ with\ a\ potential\
counterpart}\over{N_{\rm simulated\ galaxies}}}{N_{\rm galaxies\ in\
survey\ region}}
\end{equation}

We note that galaxies with low fluxes or large extents often have
higher positional errors than bright and/or compact galaxies; the
positional uncertainty of a given source is roughly the gridded Parkes
beam, 15.5$^\prime$, divided by the signal-to-noise.  The
probabilities given by these simulations are for the entire sample;
the above effect will cause slight variations among individual
galaxies. The simulations were run separately for the two regions of
the survey as the density and therefore reliability of counterparts is
field-dependent.  The difference between simulations with different
degrees of clustering around the real HI sample was used to evaluate
the approximate errors on the estimated counterpart likelihoods, $P$,
$\sim\pm$10\%.

We present in Table B1 the results of the counterpart likelihood
simulations.  The percentage of counterparts found via the automated
search that were still considered likely after a visual verification
is roughly consistent with the expected percentages calculated by
means of the simulations. The number of likely counterparts to the
galaxies in the $l=196^{\circ}$ to $212^{\circ}$ region is
substantially higher than that of the galaxies in the $l=36^{\circ}$
to $52^{\circ}$ region. The difference between the number of potential
counterparts in the two survey regions can be attributed to the lower
extinction in the latter field.  Using the 100~$\mu$m
DIRBE/\textit{IRAS} data (Schlegel et al. 1998), we calculate mean
B-band extinctions of $12^{\rm m}\!\!.0$ and $3^{\rm m}\!\!.2$ over
the $l=36^{\circ}$ to $52^{\circ}$ and $l=196^{\circ}$ to
$212^{\circ}$ regions, respectively.  This point is further
illustrated by the E(B-V) contours plotted on Figure 8. Although this
data are not well calibrated in the Galactic plane, they are still
useful for general comparison purposes.

\begin{deluxetable}{lccc}[b!]
\tabletypesize{\scriptsize}
\tablewidth{0pt}
\tablecaption {Results of Automated Search for Multiwavelength Counterparts}

\tablehead{
\colhead{}          &
\colhead{Optical}   &
\colhead{2MASS}     &
\colhead{IRAS}
}

\startdata
$l=36^{\circ}$ to $52^{\circ}$, $\Delta$r $\le 5^{\prime}$:                            &           &          &         \\
N$_{\rm counterparts}$ found via automated search                                      & 2         & 6        & 24      \\
\% expected to be real                                                                 & 5\%       & 77\%     & 18\%    \\
N$_{\rm counterparts}$ that remain after visual inspection\tablenotemark{a}            & 1(50\%)   & 4(67\%)  & 2(8\%)  \\
\tableline
$l=36^{\circ}$ to $52^{\circ}$, $\Delta$r $\le 3^{\prime}$:                            &           &          &         \\
N$_{\rm counterparts}$ found via automated search with $\Delta$r $\le 3^{\prime}$      & 1         & 5        & 14      \\
\% expected to be real                                                                 & 26\%      & 88\%     & 35\%    \\
N$_{\rm counterparts}$ that remain after visual inspection\tablenotemark{a}            & 1(100\%)  & 3(60\%)  & 2(14\%) \\
\tableline
$l=196^{\circ}$ to $212^{\circ}$, $\Delta$r $\le 5^{\prime}$:                          &           &          &         \\
N$_{\rm counterparts}$ found via automated search                                      & 17        & 31       & 15      \\
\% expected to be real                                                                 & 96\%      & 59\%     & 52\%    \\
N$_{\rm counterparts}$ that remain after visual inspection\tablenotemark{a}            & 17(100\%) & 16(52\%) & 9(60\%) \\ 
\tableline
$l=196^{\circ}$ to $212^{\circ}$, $\Delta$r $\le 3^{\prime}$:                          &           &          &         \\
N$_{\rm counterparts}$ found via automated search with $\Delta$r $\le 3^{\prime}$      & 14        & 26       & 13      \\
\% expected to be real                                                                 & 96\%      & 79\%     & 78\%    \\
N$_{\rm counterparts}$ that remain after visual inspection\tablenotemark{a}            & 14(100\%) & 14(54\%) & 9(69\%) \\ 
\enddata

\tablenotetext{a}{\#(\%) gives the number and percentage of
counterparts found via the automated search that remain on our list of
likely counterparts (and are therefore in Table 3) following a visual
inspection of all possible counterparts. In seven cases, the most
likely 2MASS counterpart was not the nearest 2MASS counterpart to the
HI position, and therefore was not the one found during our automated
search. DSS galaxies were also not found via the automated search, nor
was PGC 3097165.}

\end{deluxetable}
\end{appendix}


\newpage
\LongTables
\begin{deluxetable}{llllrrlllcc}
\tablenum{1}
\tabletypesize{\tiny}
\tablewidth{0pt}
\tablecaption {HI and Derived Properties}
\tablehead{
\colhead{}                                    &
\colhead{}                                    &
\colhead{}                                    &
\colhead{}                                    &
\colhead{}                                    &
\colhead{}                                    &
\colhead{}                                    &
\colhead{50\% Vel.}                           &
\colhead{20\% Vel.}                           &
\colhead{LG}                                  &
\colhead{Log}                                 \\
\colhead{HIZOA}                               & 
\colhead{$\alpha_{2000}$}                     &
\colhead{$\delta_{2000}$}                     &
\colhead{$\rm \ell$}                          &
\colhead{$b$}                                 &
\colhead{Flux\tablenotemark{a}}               &
\colhead{V (cz)}                              &
\colhead{Width}                               &
\colhead{Width}                               &
\colhead{Distance}                            &
\colhead{H$\rm I$ Mass}                       \\
\colhead{}                                    & 
\colhead{}                                    &
\colhead{}                                    &
\colhead{}                                    &
\colhead{}                                    &
\colhead{(Jy~km~s$^{-1}$)}                      &
\colhead{(km~s$^{-1}$)}                         &
\colhead{(km~s$^{-1}$)}                         &
\colhead{(km~s$^{-1}$)}                         &
\colhead{(Mpc)}                                 &
\colhead{($\msun$)}                 
}

\startdata
J0608+13  &   06 08 30  &   13 05 34  &   196.38  &   -3.34  &    2.27$\pm 0.14$                   &   5645$\pm    4$  & \phn50$\pm   8$  & \phn65$\pm   4$  &     74  &    9.5  \\ 
J0614+12  &   06 14 60  &   12 31 06  &   197.65  &   -2.22  &    4.03$\pm 0.22$                   &   5553$\pm    2$  &    217$\pm   7$  &    237$\pm   8$  &     73  &    9.7  \\ 
J0615+11  &   06 15 40  &   11 09 54  &   198.92  &   -2.72  &    7.67$\pm 0.28$                   &   5446$\pm    5$  &    184$\pm  14$  &    238$\pm  12$  &     71  &   10.0  \\ 
J0621+11  &   06 21 34  &   11 09 18  &   199.61  &   -1.45  &    7.74$\pm 0.20$\tablenotemark{b}  &   5599$\pm    2$  &    188$\pm   5$  &    226$\pm   7$  &     73  &   10.0  \\ 
J0621+10  &   06 21 44  &   10 25 39  &   200.27  &   -1.75  &    6.74$\pm 0.27$                   &   7467$\pm    6$  &    173$\pm  12$  &    265$\pm  40$  &     98  &   10.2  \\ 
J0622+06  &   06 22 04  &   06 29 49  &   203.79  &   -3.52  &    8.34$\pm 0.64$                   &   3598$\pm    7$  &    296$\pm  17$  &    314$\pm  23$  &     46  &    9.6  \\ 
J0622+11A &   06 22 05  &   11 22 22  &   199.48  &   -1.23  &    2.43$\pm 0.64$\tablenotemark{b}  &   5669$\pm    7$  & \phn41$\pm  17$  & \phn66$\pm  23$  &     74  &    9.5  \\ 
J0622+04  &   06 22 29  &   04 31 57  &   205.58  &   -4.34  &    7.74$\pm 0.22$                   &   2958$\pm    3$  &    306$\pm  12$  &    367$\pm  15$  &     38  &    9.4  \\ 
J0622+11B &   06 22 59  &   11 08 09  &   199.79  &   -1.15  &   19.08$\pm 0.24$\tablenotemark{b}  &   5493$\pm    1$  &    366$\pm   2$  &    409$\pm   4$  &     72  &   10.4  \\ 
J0623+14  &   06 23 19  &   14 43 41  &   196.66  &    0.60  &    5.44$\pm 1.02$                   &   5308$\pm   17$  &    286$\pm  44$  &    330$\pm  84$  &     70  &    9.8  \\ 
J0623+09  &   06 23 27  &   09 29 46  &   201.29  &   -1.82  &    6.27$\pm 0.27$                   &   8151$\pm   10$  &    238$\pm  35$  &    343$\pm  20$  &    107  &   10.2  \\ 
J0623+04  &   06 23 36  &   04 22 55  &   205.84  &   -4.17  &    3.30$\pm 0.19$                   &   2870$\pm    3$  & \phn97$\pm   9$  &    130$\pm  10$  &     37  &    9.0  \\ 
J0625+03  &   06 25 41  &   03 48 53  &   206.59  &   -3.97  &    4.80$\pm 0.22$                   &   6219$\pm    3$  &    234$\pm  11$  &    324$\pm  41$  &     81  &    9.9  \\ 
J0626+05  &   06 26 08  &   05 01 09  &   205.57  &   -3.31  &    2.67$\pm 0.11$                   &   3595$\pm    5$  &    103$\pm   7$  &    149$\pm   7$  &     46  &    9.1  \\ 
J0630+16  &   06 30 05  &   16 50 37  &   195.54  &    3.03  &   30.26$\pm 0.62$                   &   2532$\pm    3$  &    258$\pm   8$  &    328$\pm   9$  &     33  &    9.9  \\ 
J0630+08  &   06 30 09  &   08 22 37  &   203.06  &   -0.87  &    4.92$\pm 0.29$                   &\phn367$\pm    1$  & \phn33$\pm   4$  & \phn55$\pm   5$  &      3  &    7.0  \\ 
J0630+02  &   06 30 42  &   02 40 38  &   208.18  &   -3.38  &    3.59$\pm 0.13$                   &   2774$\pm    3$  &    120$\pm   6$  &    144$\pm   5$  &     35  &    9.0  \\ 
J0632+06  &   06 32 53  &   06 17 16  &   205.22  &   -1.24  &    3.28$\pm 0.21$                   &   3147$\pm    6$  &    117$\pm  19$  &    168$\pm  10$  &     40  &    9.1  \\ 
J0633+10  &   06 33 59  &   10 40 40  &   201.45  &    1.03  &    1.68$\pm 0.09$                   &   7547$\pm    2$  & \phn43$\pm   3$  & \phn65$\pm   6$  &     99  &    9.6  \\ 
J0634+09  &   06 34 05  &   09 13 17  &   202.76  &    0.38  &    1.29$\pm 0.10$                   &   3876$\pm    1$  & \phn37$\pm   5$  & \phn48$\pm   5$  &     50  &    8.9  \\ 
J0634+00  &   06 34 20  &   00 03 37  &   210.93  &   -3.78  &    1.88$\pm 0.11$                   &   6183$\pm    2$  & \phn56$\pm   5$  & \phn86$\pm   7$  &     80  &    9.5  \\ 
J0635+14A &   06 35 17  &   14 59 15  &   197.77  &    3.29  &    6.44$\pm 0.15$                   &   3827$\pm    4$  &    275$\pm  11$  &    328$\pm  11$  &     50  &    9.6  \\ 
J0635+02  &   06 35 44  &   02 24 17  &   209.00  &   -2.39  &    4.07$\pm 0.33$                   &   6334$\pm   14$  & \phn41$\pm  23$  &    228$\pm  41$  &     83  &    9.8  \\ 
J0635+03  &   06 35 51  &   03 09 45  &   208.34  &   -2.02  &    1.49$\pm 0.19$                   &   2832$\pm    3$  & \phn33$\pm   8$  & \phn59$\pm  12$  &     36  &    8.7  \\ 
J0635+11  &   06 35 54  &   11 09 44  &   201.24  &    1.67  &    9.11$\pm 0.30$                   &   3734$\pm    5$  &    408$\pm   9$  &    479$\pm   5$  &     48  &    9.7  \\ 
J0635+14B &   06 35 54  &   14 36 21  &   198.18  &    3.25  &   16.47$\pm 0.13$                   &   4020$\pm    1$  &    329$\pm   2$  &    360$\pm   2$  &     52  &   10.0  \\ 
J0636+00  &   06 36 24  &   00 56 42  &   210.37  &   -2.91  &   17.39$\pm 0.16$                   &   2723$\pm    1$  &    144$\pm   2$  &    188$\pm   3$  &     34  &    9.7  \\ 
J0636+04  &   06 36 51  &   04 03 60  &   207.65  &   -1.38  &    5.45$\pm 0.24$                   &   3520$\pm    4$  &    185$\pm  13$  &    196$\pm  22$  &     45  &    9.4  \\ 
J0637+03  &   06 37 39  &   03 24 13  &   208.33  &   -1.51  &    5.93$\pm 0.35$                   &   3431$\pm    3$  &    160$\pm   7$  &    172$\pm   7$  &     44  &    9.4  \\ 
J0641+01  &   06 41 03  &   01 50 60  &   210.10  &   -1.47  &    6.78$\pm 0.19$                   &   2795$\pm    2$  &    181$\pm   5$  &    199$\pm   4$  &     35  &    9.3  \\ 
J0643+06  &   06 43 18  &   06 57 57  &   205.81  &    1.37  &    7.75$\pm 0.32$                   &   6903$\pm    3$  &    356$\pm   9$  &    381$\pm  12$  &     90  &   10.2  \\ 
J0644+12  &   06 44 01  &   12 25 58  &   201.01  &    4.02  &    9.34$\pm 0.26$                   &   3925$\pm    2$  &    403$\pm   6$  &    425$\pm  15$  &     51  &    9.8  \\ 
J0649+09  &   06 49 35  &   09 43 03  &   204.06  &    4.01  &   15.92$\pm 0.30$                   &   7513$\pm    3$  &    268$\pm  15$  &    302$\pm   8$  &     99  &   10.6  \\ 
J0653+07  &   06 53 42  &   07 09 54  &   206.80  &    3.76  &    5.63$\pm 0.24$                   &   7035$\pm    4$  &    329$\pm   6$  &    360$\pm  32$  &     92  &   10.1  \\ 
J0654+08  &   06 54 10  &   08 35 36  &   205.58  &    4.51  &    5.79$\pm 0.15$                   &   2625$\pm    4$  &    128$\pm   9$  &    172$\pm   9$  &     33  &    9.2  \\ 
J0656+06A &   06 56 26  &   06 01 44  &   208.13  &    3.85  &   12.29$\pm 0.21$\tablenotemark{b}  &   6391$\pm    3$  &    184$\pm   6$  &    228$\pm   4$  &     83  &   10.3  \\ 
J0656+06B &   06 56 28  &   06 16 43  &   207.91  &    3.97  &   10.69$\pm 0.21$\tablenotemark{b}  &   6784$\pm    5$  &    207$\pm   9$  &    286$\pm  14$  &     89  &   10.3  \\ 
J0659+06  &   06 59 37  &   06 26 15  &   208.12  &    4.74  &    4.38$\pm 0.18$                   &   6365$\pm    3$  &    145$\pm   7$  &    168$\pm   8$  &     83  &    9.9  \\ 
J0700+01  &   07 00 58  &   01 56 04  &   212.30  &    3.00  &   15.06$\pm 0.32$                   &   1762$\pm    1$  &    334$\pm   3$  &    360$\pm   3$  &     21  &    9.2  \\ 
J0702+04  &   07 02 24  &   04 52 39  &   209.83  &    4.66  &    7.23$\pm 0.23$                   &   6934$\pm    3$  &    241$\pm   5$  &    263$\pm   6$  &     90  &   10.1  \\ 
J0702+03  &   07 02 50  &   03 13 37  &   211.35  &    4.00  &    4.20$\pm 0.26$                   &   3551$\pm    3$  & \phn86$\pm  10$  &    121$\pm  20$  &     45  &    9.3  \\ 
J0705+02  &   07 05 39  &   02 37 37  &   212.21  &    4.36  &    5.17$\pm 0.15$                   &   1745$\pm    1$  & \phn34$\pm   2$  & \phn49$\pm   2$  &     21  &    8.7  \\ 
J1843+06  &   18 43 42  &   06 33 59  &    37.91  &    4.70  &    6.34$\pm 0.47$                   &   8918$\pm    6$  &    266$\pm  18$  &    304$\pm  31$  &    121  &   10.3  \\ 
J1852+10  &   18 52 59  &   10 25 49  &    42.41  &    4.40  &    5.17$\pm 0.16$                   &   4906$\pm    4$  &    238$\pm  12$  &    275$\pm   7$  &     68  &    9.8  \\ 
J1853+09  &   18 53 49  &   09 51 11  &    41.98  &    3.95  &   18.30$\pm 0.27$                   &   4729$\pm    1$  &    320$\pm   3$  &    350$\pm   3$  &     66  &   10.3  \\ 
J1855+07  &   18 55 22  &   07 38 34  &    40.18  &    2.61  &    6.08$\pm 0.42$                   &   6159$\pm    4$  &    266$\pm  11$  &    330$\pm  36$  &     85  &   10.0  \\ 
J1857+13  &   18 57 22  &   13 26 26  &    45.59  &    4.80  &    5.01$\pm 0.30$                   &   8095$\pm    5$  &    173$\pm  20$  &\nodata\tablenotemark{c}&    111  &   10.2  \\ 
J1900+13  &   19 00 06  &   13 33 21  &    45.99  &    4.25  &    9.40$\pm 0.24$                   &   4658$\pm    2$  &    234$\pm   4$  &    252$\pm   7$  &     65  &   10.0  \\ 
J1901+06  &   19 01 40  &   06 51 44  &    40.20  &    0.86  &   17.27$\pm 0.17$                   &   2945$\pm    1$  & \phn61$\pm   2$  & \phn88$\pm   1$  &     42  &    9.9  \\ 
J1904+03A &   19 04 15  &   03 06 47  &    37.16  &   -1.43  &    5.60$\pm 0.22$\tablenotemark{b}  &   3289$\pm    2$  & \phn76$\pm   5$  &    125$\pm   9$  &     46  &    9.5  \\ 
J1904+03B &   19 04 27  &   03 11 32  &    37.26  &   -1.43  &    5.53$\pm 0.33$\tablenotemark{b}  &   3195$\pm    4$  &    232$\pm  15$  &    260$\pm   9$  &     45  &    9.4  \\ 
J1906+12  &   19 06 13  &   12 57 17  &    46.13  &    2.66  &    8.47$\pm 0.36$                   &   2730$\pm    2$  &    267$\pm   6$  &    293$\pm  20$  &     39  &    9.5  \\ 
J1906+07  &   19 06 53  &   07 33 52  &    41.42  &    0.04  &    5.16$\pm 0.30$                   &   3095$\pm    5$  &    138$\pm  12$  &    188$\pm  18$  &     44  &    9.4  \\ 
J1907+14  &   19 07 33  &   14 02 54  &    47.26  &    2.87  &    4.18$\pm 0.29$                   &   7290$\pm    3$  & \phn89$\pm  10$  &\nodata\tablenotemark{c}&    100  &   10.0  \\ 
J1908+05  &   19 08 26  &   05 59 34  &    40.20  &   -1.03  &    5.97$\pm 0.28$                   &   4567$\pm    4$  &    168$\pm   9$  &    196$\pm  14$  &     63  &    9.8  \\ 
J1910+00  &   19 10 25  &   00 33 28  &    35.60  &   -3.97  &   16.60$\pm 0.36$                   &   1494$\pm    1$  &    180$\pm   5$  &    194$\pm   4$  &     22  &    9.3  \\ 
J1912+13  &   19 12 39  &   13 24 33  &    47.26  &    1.48  &    7.09$\pm 0.15$                   &   2774$\pm    1$  &    102$\pm   3$  &    117$\pm   2$  &     40  &    9.4  \\ 
J1912+02  &   19 12 45  &   02 56 32  &    37.99  &   -3.39  &   15.23$\pm 0.42$                   &   6541$\pm   10$  &    470$\pm  22$  &    597$\pm  20$  &     90  &   10.5  \\ 
J1913+17  &   19 13 49  &   17 35 31  &    51.10  &    3.16  &    2.58$\pm 0.20$                   &   4726$\pm    4$  & \phn80$\pm  11$  &    124$\pm   9$  &     66  &    9.4  \\ 
J1913+16  &   19 13 59  &   16 54 60  &    50.52  &    2.81  &   11.47$\pm 0.67$                   &   6263$\pm    4$  &    539$\pm   9$  &    616$\pm  22$  &     87  &   10.3  \\ 
J1914+10  &   19 14 57  &   10 17 47  &    44.76  &   -0.47  &   20.66$\pm 0.37$                   &\phn654$\pm    0$  & \phn66$\pm   2$  & \phn83$\pm   2$  &     12  &    8.8  \\ 
J1917+07  &   19 17 27  &   07 49 14  &    42.86  &   -2.17  &    6.89$\pm 0.19$                   &   3033$\pm    2$  &    226$\pm   6$  &    243$\pm   6$  &     43  &    9.5  \\ 
J1917+04  &   19 17 38  &   04 26 33  &    39.89  &   -3.78  &   11.18$\pm 0.33$                   &   6330$\pm   10$  &    325$\pm  22$  &    416$\pm  11$  &     87  &   10.3  \\ 
J1918+16  &   19 18 38  &   16 10 13  &    50.38  &    1.48  &    7.08$\pm 0.35$                   &   6639$\pm   13$  &    165$\pm  37$  &    279$\pm  17$  &     92  &   10.1  \\ 
J1919+14  &   19 19 59  &   14 04 01  &    48.67  &    0.20  &   13.70$\pm 0.44$                   &   2809$\pm    2$  &    137$\pm   7$  &    195$\pm   8$  &     40  &    9.7  \\ 
J1921+14  &   19 21 36  &   14 54 26  &    49.60  &    0.25  &    6.56$\pm 0.22$                   &   4080$\pm    2$  & \phn72$\pm   5$  & \phn88$\pm   6$  &     57  &    9.7  \\ 
J1921+08  &   19 21 59  &   08 16 59  &    43.79  &   -2.94  &    5.97$\pm 0.39$                   &   3115$\pm    4$  &    100$\pm  10$  &    125$\pm  15$  &     44  &    9.4  \\ 
J1926+08  &   19 26 02  &   08 16 09  &    44.25  &   -3.84  &    2.62$\pm 0.18$                   &   3088$\pm    2$  & \phn67$\pm  11$  &    113$\pm  15$  &     44  &    9.1  \\ 
J1927+12  &   19 27 36  &   12 19 20  &    48.01  &   -2.25  &    5.38$\pm 0.38$                   &   8604$\pm    6$  &    138$\pm  17$  &    204$\pm  15$  &    118  &   10.2  \\ 
J1927+09  &   19 27 57  &   09 26 51  &    45.52  &   -3.70  &    1.69$\pm 0.22$                   &   7920$\pm    4$  & \phn48$\pm   7$  & \phn77$\pm  24$  &    108  &    9.7  \\ 
J1929+08  &   19 29 18  &   08 03 55  &    44.46  &   -4.64  &   19.08$\pm 0.97$                   &   3097$\pm    6$  &    195$\pm  13$  &    341$\pm  14$  &     44  &    9.9  \\ 
J1929+11  &   19 29 24  &   11 36 44  &    47.59  &   -2.98  &    5.78$\pm 0.31$                   &   6331$\pm    6$  &    351$\pm  16$  &    388$\pm  12$  &     87  &   10.0  \\ 
J1930+12  &   19 30 29  &   12 12 07  &    48.24  &   -2.93  &    6.07$\pm 0.28$                   &   6684$\pm    3$  &    213$\pm  11$  &    240$\pm  13$  &     92  &   10.1  \\ 
J1930+11  &   19 30 36  &   11 18 53  &    47.47  &   -3.38  &    3.11$\pm 0.16$                   &   3174$\pm    3$  &    108$\pm   7$  &    124$\pm  11$  &     45  &    9.2  \\ 
J1933+10  &   19 33 56  &   10 42 35  &    47.33  &   -4.39  &    6.67$\pm 0.35$                   &   5255$\pm    3$  &    240$\pm   8$  &    282$\pm   9$  &     73  &    9.9  \\ 
J1937+14  &   19 37 12  &   14 42 55  &    51.23  &   -3.15  &    3.20$\pm 0.15$                   &   4485$\pm    2$  & \phn59$\pm   4$  & \phn74$\pm   5$  &     63  &    9.5  \\ 
J1940+11  &   19 40 36  &   11 54 18  &    49.18  &   -5.24  &    1.49$\pm 0.17$                   &\phn580$\pm    2$  & \phn20$\pm   5$  & \phn32$\pm  11$  &     11  &    7.6  \\ 
\enddata

\tablenotetext{a}{The error we estimate for the integrated flux does not take into
account the uncertainty due to baseline subtraction or the uncertainty
in calibration; the total error on the integrated flux is likely to
be $\sim$ 10-15\%}

\tablenotetext{b}{The flux was calculated over a smaller region to isolate the galaxy 
emission from that arising from a nearby galaxy. This may have led to
an underestimation of the total flux.}

\tablenotetext{c}{The 20\% velocity width could not be accurately measured.}

\end{deluxetable}


\clearpage
\begin{landscape}
\begin{deluxetable}{lccccccclclc}
\pagestyle{empty}
\hspace{-4.cm}
\tablenum{2}
\tabletypesize{\tiny}
\tablewidth{0pt}
\tablecaption {Multiwavelength Counterparts}
\tablehead{
\colhead{Name}                                    &
\colhead{b}                                       &
\colhead{E(B-V)\tablenotemark{a}}                 &
\colhead{HI Counterpart}                          &
\colhead{Sep}                                     &
\colhead{$\Delta$v\tablenotemark{b}}              &
\colhead{Optical Counterpart\tablenotemark{c}}                     &
\colhead{Sep}                                     &
\colhead{2MASS Counterpart\tablenotemark{c}}                       &
\colhead{Sep}                                     &
\colhead{IRAS Counterpart\tablenotemark{c}}                        &
\colhead{Sep}                                     \\
\colhead{}                                        &
\colhead{}                                        &
\colhead{}                                        &
\colhead{}                                        &
\colhead{($^{\prime}$)}                           &
\colhead{(km~s$^{-1}$)}                           &
\colhead{}                                        &
\colhead{($^{\prime}$)}                           &
\colhead{(2MASX)}                                 &
\colhead{($^{\prime}$)}                           &
\colhead{(IRAS)}                                  &
\colhead{($^{\prime}$)}                           
}

\startdata 
  J0615+11                   &  -2.72  & 0.8     &  ADBS J061543+1110        & 1.4         & \ \ \ 9.4     & DSS galaxy                    & \nodata       &  J06154496+1110227                   & 1.3      & \nodata           & \nodata        \\
  J0621+11                   &  -1.45  & 0.9     &  \nodata                  & \nodata     & \nodata       & ZOAG G199.66-01.50            & 4.2           &  J06212888+1105105                   & 4.2      & \nodata           & \nodata        \\
  J0621+10                   &  -1.75  & 0.7     &  CAP 0618+10              & 1.0         & \ \ \ 6.1     & WEIN 192                      & 1.0           &  J06214000+1024565                   & 1.0      & 06188+1026        & 1.0            \\ 
  J0622+06                   &  -3.52  & 0.5     &  \nodata                  & \nodata     & \nodata       & ZOAG G203.78-03.50            & 1.7           &  J06220859+0631063                   & 1.7      & 06194+0632        & 2.0            \\
  J0622+11A                  &  -1.23  & 0.9     &  \nodata                  & \nodata     & \nodata       & DSS galaxy                    & 4.0           &  \nodata                             & \nodata  & \nodata           & \nodata        \\
  J0622+04                   &  -4.34  & 0.6     &  CAP 0619+04              & 0.9         & \ \ \ 4.9     & ZOAG G205.58-04.36            & 0.9           &  J06222633+0431236                   & 0.9      & 06197+0432        & 0.8            \\  
  J0622+11B                  &  -1.15  & 1.1     &  ADBS J062302+1108        & 1.1         & \ \ 50.0      & DSS galaxy                    & \nodata       &  J06225815+1108312                   & 0.4      & 06202+1109        & 0.9            \\
  J0623+14                   &   0.60  & 0.9     &  WEIN 179                 & 1.7         & \ \ 54.3      & WEIN 179                      & 1.7           &  J06231132+1443129                   & 1.7      & \nodata           & \nodata        \\
  J0623+09                   &  -1.82  & 0.6     &  \nodata                  & \nodata     & \nodata       & DSS galaxy                    & \nodata       &  J06233278+0930182                   & 1.8      & \nodata           & \nodata        \\
  J0625+03                   &  -3.97  & 0.6     &  \nodata                  & \nodata     & \nodata       & see note\tablenotemark{d}     & \nodata       &  see note\tablenotemark{d}           &\nodata   & \nodata           & \nodata        \\  
  J0630+16                   &   3.03  & 0.6     &  \nodata                  & \nodata     & \nodata       & ZOAG G195.62+03.04            & 4.7           &  J06301575+1646422                   & 4.7      & \nodata           & \nodata        \\
  J0630+02                   &  -3.38  & 0.9     &  \nodata                  & \nodata     & \nodata       & DSS galaxy                    & \nodata       &  J06303567+0238356\tablenotemark{e}  & 2.4      & \nodata           & \nodata        \\
  J0633+10                   &  1.03   & 1.5     &  \nodata                  & \nodata     & \nodata       & \nodata                       & \nodata       &  J06334997+1039414                   & 2.4      & \nodata           & \nodata        \\
  J0634+09                   &  0.38   & 0.8     &  \nodata                  & \nodata     & \nodata       & DSS galaxy                    & 0.4           &  \nodata                             & \nodata  & \nodata           & \nodata        \\                  
  J0635+14A                  &  3.29   & 0.5     &  UGC 03498                & 1.0         & \ -27.3       & UGC 03498                     & 1.0           &  J06352031+1458436                   & 0.9      & 06324+1501        & 1.0            \\
  J0635+02                   & -2.39   & 1.8     &  \nodata                  & \nodata     & \nodata       & DSS galaxy                    & \nodata       &  J06353276+0224228                   & 2.6      & \nodata           & \nodata        \\
  J0635+11\tablenotemark{f}  &  1.67   & 1.5     &  ADBS J063549+1107        & 2.4         & \ \ 31.3      & DSS galaxy                    & 1.6           &  \nodata                             & \nodata  & \nodata           & \nodata        \\
                             & \nodata & \nodata &  ADBS J063606+1109        & 2.4         & -163.8        & \nodata                       & \nodata       &  \nodata                             & \nodata  & \nodata           & \nodata        \\
  J0635+14B                  &  3.25   & 0.6     &  \nodata                  & \nodata     & \nodata       & PGC 3097165                   & 0.8           &  J06355671+1435584                   & 0.7      & \nodata           & \nodata        \\
  J0636+00                   & -2.91   & 1.5     &  \nodata                  & \nodata     & \nodata       & ZOAG G210.39-02.92            & 1.0           &  \nodata                             & \nodata  & 06338+0058        & 0.9            \\
  J0643+06                   &  1.37   & 0.8     &  WEIN 206                 & 1.4         & \ \ 2.4       & WEIN 206                      & 1.4           &  J06431155+0657324                   & 1.4      & \nodata           & \nodata        \\
  J0644+12                   &  4.02   & 0.4     &  UGC 03524                & 1.9         & \ \ -2.9      & UGC 03524                     & 1.9           &  J06440058+1224067                   & 1.9      & 06412+1227        & 1.8            \\
  J0649+09                   &  4.01   & 0.3     &  \nodata                  & \nodata     & \nodata       & ZOAG G204.10+03.97            & 3.4           &  J06493148+0939437                   & 3.4      & \nodata           & \nodata        \\
  J0653+07                   &  3.76   & 0.3     &  \nodata                  & \nodata     & \nodata       & ZOAG G206.80+03.71            & 2.8           &  J06533152+0708263                   & 2.8      & \nodata           & \nodata        \\
  J0654+08                   &  4.51   & 0.2     &  ADBS J065406+0834        & 1.4         & \ \ 13.9      & ZOAG G205.59+04.48            & 1.4           &  \nodata                             & \nodata  & \nodata           & \nodata        \\
  J0656+06A                  &  3.85   & 0.4     &  CAP 0653+06              & 1.9         & \ \ -0.3      & ZOAG G208.14+03.82            & 1.9           &  J06562110+0600057                   & 1.9      & \nodata           & \nodata        \\
  J0656+06B                  &  3.97   & 0.3     &  UGC 03607                & 2.4         & \ -18.0       & UGC 03607                     & 2.4           &  J06561786+0616066                   & 2.4      & 06536+0620        & 2.3            \\
  J0659+06                   &  4.74   & 0.3     &  \nodata                  & \nodata     & \nodata       & DSS galaxy\tablenotemark{g}   & 1.1           &  \nodata                             & \nodata  & \nodata           & \nodata        \\
  J0700+01                   &  3.00   & 0.5     &  UGC 03630                & 1.9         & \ \ -0.4      & UGC 03630                     & 1.9           &  J07010328+0154406                   & 1.9      & 06584+0158        & 2.0            \\
  J0702+04                   &  4.66   & 0.3     &  CAP 0659+04              & 1.6         & \ \ -3.9      & ZOAG G209.84+04.63            & 1.6           &  J07021916+0451191                   & 1.6      & \nodata           & \nodata        \\
  J0702+03                   &  4.00   & 0.4     &  \nodata                  & \nodata     & \nodata       & DSS galaxy                    & 2.4           &  \nodata                             & \nodata  & \nodata           & \nodata        \\ 
  J0705+02                   &  4.36   & 0.5     &  [H92] 16                 & 0.3         & \ \ \ 0.3     & \nodata                       & \nodata       &  \nodata                             & \nodata  & \nodata           & \nodata        \\
  J1857+13                   &  4.80   & 0.9     &  \nodata                  & \nodata     & \nodata       & DSS galaxy                    & \nodata       &  J18573802+1328565                   & 4.8      & \nodata           & \nodata        \\
  J1901+06                   &  0.86   & 5.6     &  Dw040.0+0.9              & 1.3         & \ -10.5       & \nodata                       & \nodata       &  \nodata                             & \nodata  & \nodata           & \nodata        \\
  J1906+12                   &  2.66   & 1.5     &  2MASX J19061820+1256195  & 1.6         & \ \ -8.4      & DSS galaxy                    & \nodata       &  J19061820+1256195                   & 1.6      & \nodata           & \nodata        \\
  J1910+00                   & -3.97   & 0.7     &  HIZSS 110                & 3.0         & \ \ \ 9.0     & \nodata                       & \nodata       &  \nodata                             & \nodata  & \nodata           & \nodata        \\
  J1913+16                   &  2.81   & 2.4     &  \nodata                  & \nodata     & \nodata       & DSS galaxy                    & \nodata       &  J19135226+1655279                   & 1.5      & \nodata           & \nodata        \\
  J1914+10                   & -0.47   & 11.7    &  Dw044.8-0.5              & 0.3         & \ \ 15.5      & \nodata                       & \nodata       &  \nodata                             & \nodata  & \nodata           & \nodata        \\
  J1917+07                   & -2.17   & 1.9     &  \nodata                  & \nodata     & \nodata       & DSS galaxy                    & 0.5           &  \nodata                             & \nodata  & \nodata           & \nodata        \\
  J1917+04                   & -3.78   & 1.0     &  ZOAG G039.88-03.78       & 0.1         & \ \ \ 7.2     & ZOAG G039.88-03.78            & 0.1           &  J19173769+0426248                   & 0.1      & 19151+0421        & 0.2            \\
  J1929+08                   & -4.64   & 0.5     &  \nodata                  & \nodata     & \nodata       & DSS galaxy                    & 1.3           &  \nodata                             & \nodata  & 19269+0756        & 1.3            \\ 
  J1933+10                   & -4.39   & 0.5     &  \nodata                  & \nodata     & \nodata       & DSS galaxy                    & \nodata       &  J19335171+1042247                   & 0.8      & \nodata           & \nodata        \\
\enddata

\tablenotetext{a}{from DIRBE/\textit{IRAS} 100 $\mu$m extinction maps (Schlegel
et al. 1998)}

\tablenotetext{b}{$\Delta v = v_{\rm counterpart} - v$ (km~s$^{-1}$)}

\tablenotetext{c}{The references for counterparts are as follows:
(ADBS) Rosenberg \& Schneider 2000; (CAP) Pantoja et al. 1997; (Dw)
Henning et al. 1998, Rivers 2000; (HIZSS) Henning et al. 2000; (PGC)
http://leda.univ-lyon1.fr/; (UGC) Nilson 1973, (WEIN) Weinberger 1980;
(ZOAG) Seeberger, Saurer, \& Weinberger 1996; (2MASX
J19061820+1256195) Jarrett et al. 2000 }

\tablenotetext{d}{There are two possible counterparts to this source,
2MASS J06254183+0346338 at a separation of 2.3$^{\prime}$ and a DSS
galaxy at $\alpha=06^{\rm h}25^{\rm m}45^{\rm s}\!\!.1$,  
$\delta=03^{\circ}47^{\prime}55^{\prime\prime}$ (separation =
1.4$^{\prime}$).  We do not have enough information to decide between
the two.}

\tablenotetext{e}{We note that there is another possible 2MASS
counterpart, J06303033+0240026, at a separation of 2.7$^{\prime}$.
This candidate appears to be of earlier type than that chosen above,
but can not be excluded.}

\tablenotetext{f}{Our HI detection was resolved into two galaxies
separated by 4$^{\prime}$ by the ADBS. We are unable to say to which
HI galaxy the counterpart belongs.}

\tablenotetext{g}{In addition to the counterpart listed here, we see
another DSS galaxy at $\alpha=06^{\rm h}59^{\rm m}38^{\rm s}\!\!.1$
and $\delta=06^{\circ}25^{\prime}19^{\prime\prime}$ as well as two
2MASS candidates: J06595034+0623322 (sep=4.3$^{\prime}$) and
J06595034+0623322 (sep=4.4$^{\prime}$).  From morphology and
separation arguments, we believe the candidate we have chosen is the
most likely, but we can not rule out those listed here.}

\end{deluxetable}
\clearpage
\end{landscape}

\end{document}